\RequirePackage{fix-cm}
\documentclass[twocolumn]{svjour3}          
\usepackage{amssymb}
\usepackage{amsmath}
\usepackage{stfloats}
\usepackage{bm}
\usepackage{graphicx}
\usepackage{float}
\usepackage{caption}
\usepackage{lineno}
\usepackage[numbers,sort&compress]{natbib}
\usepackage{algorithm}
\usepackage{algpseudocode}
\usepackage{yhmath}
\usepackage[colorlinks=true,linkcolor=blue,citecolor=blue,anchorcolor=blue,urlcolor=blue]{hyperref}
\begin{document}

\title{Computational study on microstructure evolution and magnetic property of laser additively manufactured magnetic materials}

\titlerunning{Computational study on additive manufacturing of magnetic materials}        

\author{Min Yi \and Bai-Xiang Xu \and Oliver Gutfleisch}

\authorrunning{M. Yi et al.} 

\institute{M. Yi , B.-X. Xu, O. Gutfleisch \at
              Institute of Materials Science, Technische Universit\"at Darmstadt, Darmstadt 64287, Germany  \\
              Tel.: +49 6151 16-22922\\
              Fax: +49 6151 16-21034\\
              \email{yi@mfm.tu-darmstadt.de; xu@mfm.tu-darmstadt.de}     
}
 
\date{Received: July 10, 2018 / Accepted: } 

\maketitle

\begin{abstract}
IAdditive manufacturing (AM) offers an unprecedented opportunity for the quick production of complex shaped parts directly from a powder precursor. But its application to functional materials in general and magnetic materials in particular is still at the very beginning. Here we present the first attempt to computationally study the microstructure evolution and magnetic properties of magnetic materials (e.g. Fe-Ni alloys) processed by selective laser melting (SLM). SLM process induced thermal history and thus the residual stress distribution in Fe-Ni alloys are calculated by finite element analysis (FEA). The evolution and distribution of the $\gamma$-Fe-Ni and FeNi$_3$ phase fractions were predicted by using the temperature information from FEA and the output from CALculation of PHAse Diagrams (CALPHAD). Based on the relation between residual stress and magnetoelastic energy, magnetic properties of SLM processed Fe-Ni alloys (magnetic coercivity, remanent magnetization, and magnetic domain structure) are examined by micromagnetic simulations. The calculated coercivity is found to be in line with the experimentally measured values of SLM-processed Fe-Ni alloys. This computation study demonstrates a feasible approach for the simulation of additively manufactured magnetic materials by integrating FEA, CALPHAD, and micromagnetics.
\end{abstract}

\section{Introduction}
Fe-Ni permalloys are typical soft magnetic materials with extraordinary magnetic, mechanical, and electrical properties \cite{arnold1923permalloy}. Due to their low coercivity, high magnetoconductivity, high permeability, and moderate saturation magnetization, they are of great interests for applications in electromagnetic devices, including transformers, sensors, and electric motors \cite{kwiatkowski1986permalloy,ganz1946applications,ripka2008sensors}. In order to realize these applications, suitable manufacturing techniques have to be identified since they significantly affect the magnetic properties. In the past, numerous conventional manufacturing methods such as sintering, thermal spraying, ball milling, and magnetron sputtering have been used to obtain the desirable performance of Fe-Ni alloys. Nevertheless, within the scope of these methods, the direct consolidation of different types of powders into bulk magnetic components with magnetism preserved is always challenging.
Moreover, these conventional methods may lead to the decrease of magnetic properties due to the excessive grain growth under a low-speed heating and cooling. They are also weak in producing precise magnetic components with complex shape and geometry.

Selective laser melting (SLM), as a typical additive manufacturing (AM) technique, enables the quick production of complex shaped three-dimensional (3D) parts directly from metal powders. Up to now, a large number of studies about SLM-AM or electron-beam-AM have been focused on structural materials with mechanical properties as the focus, such as aluminium alloys \cite{martin20173d}, Ti-Al-V alloys \cite{kobryn2001laser,yan2016multi,yan2017modeling}, Ni-based superalloys \cite{keller2017application}, stainless steel \cite{wang2017additively,johnson2018simulation}, etc. In contrast, the application of SLM-AM to functional materials is still in its infancy. Nonetheless, SLM-AM undeniably provides a promising route for breaking the bottlenecks of traditional techniques to fabricate complex shaped functional and miniaturized magnetic devices or systems directly from metal powders. Ongoing efforts have been devoted to the production of magnetic materials by SLM-AM.
The initial work was carried out on the SLM processing of magnetic Fe-Ni alloy by Zhang \textit{et al}. \cite{zhang2012studies, zhang2013microstructure, zhang2013magnetic}. Depending on the composition and processing, Fe-Ni alloy can be either a soft magnetic material or in the L1$_0$ phase as a rare-earth-free alternative for permanent magnets \cite{poirier2015intrinsic, bordeaux2016thermodynamic}, thus making Fe-Ni alloy a very interesting material. Later, Moore \textit{et al.} fabricated magnetocaloric La(Fe,Co,Si)$_{13}$ geometries by SLM \cite{moore2013selective}.
However, after the early work \cite{zhang2012studies,zhang2013microstructure, zhang2013magnetic,moore2013selective} in 2012 and 2013, it is found from the literature survey that few studies followed. Only lately in 2016$-$2018, studies continue with focus on SLM processed magnetic materials such as Fe-Si alloy \cite{garibaldi2018effect}, Fe-Si-Cr alloy \cite{jhong2016microstructure}, Fe-80$\%$Ni permalloy \cite{shishkovsky2016peculiarities}, Fe-30$\%$Ni alloy \cite{mikler2017tuning}, Ni-Fe-V and Ni-Fe-Mo permalloys \cite{mikler2017laser1,mikler2017laser2}, Fe-Co-1.5V soft magnetic alloy \cite{kustas2018characterization}, permanent magnets including NdFeB \cite{jacimovic2017net} and AlNiCo \cite{white2017net}, ect. Electron beam melting (EBM) is also tried to produce MnAl(C) magnets \cite{popov2018prospects}.
Apart from the SLM and EBM based AM technique, other 3D printing technologies without high energy input and high temperature, such as binder jetting and material extrusion, are recently applied to the production of polymer-bonded magnets \cite{li2016big,paranthaman2016binder,li2017additive,huber20173d,huber2017topology}.
These experimental studies reveal the notable effect of AM process on the microstructure and magnetic properties of magnetic materials, and provide insight into the challenges for the design and control of magnetic properties by AM.

Despite of these recent experimental efforts, no literature is found about the modeling and simulation of the fabrication of magnetic alloys by SLM-AM. Almost all computational studies are dedicated to the structural materials by SLM-AM  with a focus on the temperature, microstructure, residual stress, strength and ductility \cite{smith2016linking, markl2016multiscale, yan2018data, yang2018prediction}, possibly driven by the related experimental contributions which are continuously flourishing. As for fabricating magnetic alloys by SLM-AM, numerical simulations are also essential for the optimization of SLM-AM processes without intensive and expensive trial-and-error experimental iterations, as well as for the understanding of underlying physical phenomena which are difficult to observe experimentally.

In this work, taking magnetic Fe-Ni alloy as a model material, we attempt to computationally predict the microstructure evolution and coercivity of SLM processed magnetic materials through the integration of finite element analysis (FEA), CALculation of PHAse Diagrams (CALPHAD), and micromagnetic simulations. Temperature history and distribution were calculated by FEA within the framework of heat transfer. By using the temperature information as the input, thermomechanical simulation by FEA were performed to get the residual stress distribution. Furthermore, integrating temperature results with CALPHAD output resulted in the temporal evolution of liquid, $\gamma$-Fe-Ni phase, and FeNi$_3$ phase. Finally, by incorporating the residual stress into the magnetoelastic energy of micromagnetics, the magnetic hysteresis and coercivity were calculated.
It is anticipated the computational study could provide a possible general routine or procedure for enlarging the process understanding of the underlying physical mechanisms in the SLM processed magnetic materials.
\begin{figure}[!b]
\centering
\includegraphics[width=5.5cm]{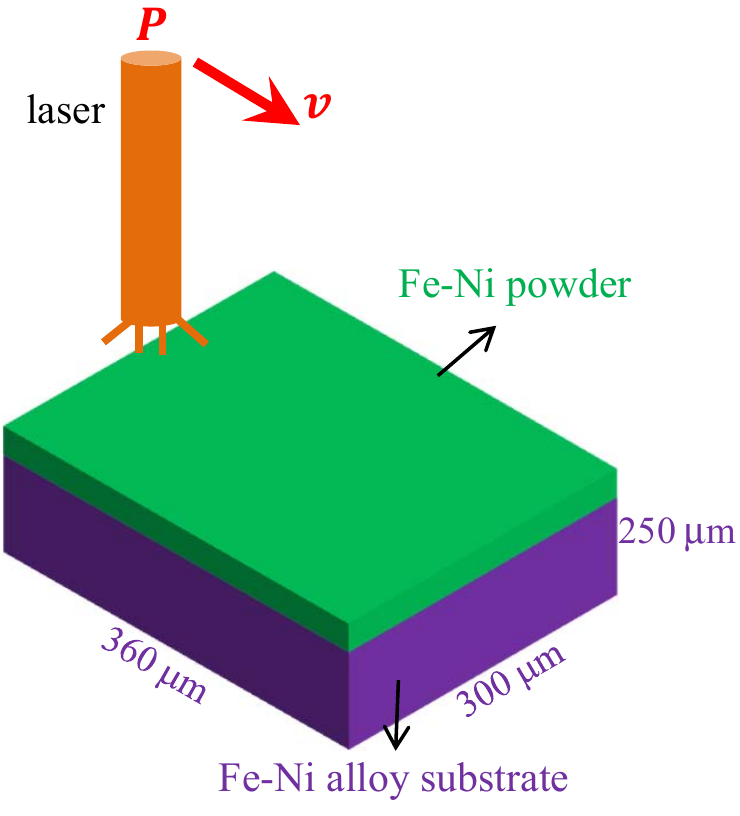}
\caption{Schematics of fabricating magnetic Fe-Ni alloys by SLM process.}
\label{f1}
\end{figure}

\begin{figure*}[!t]
\centering
\includegraphics[width=13.4cm]{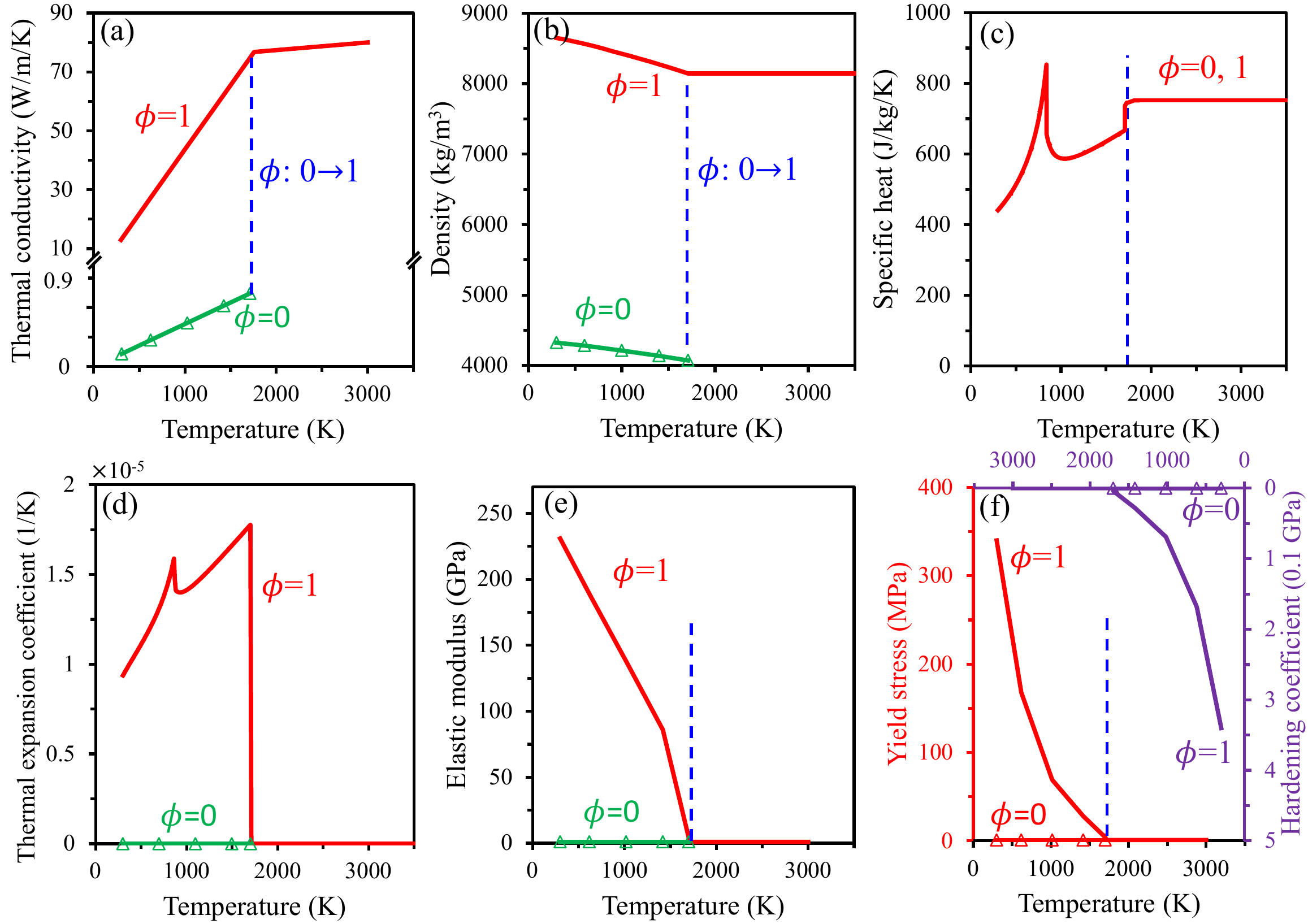}
\caption{Material parameters as a function of temperature $T$ and material state $\phi$. The blue dashed vertical line represents the temperature $T_s \approx T_l=1709$ K.}
\label{f2}
\end{figure*}
\section{Thermal analysis}
The fabrication of magnetic Fe-Ni alloy by direct SLM processing of powders is illustrated in Fig. \ref{f1}. An Fe-Ni alloy substrate with a dimension of 360 $\mu$m $\times$ 300 $\mu$m $\times$ 250 $\mu$m is chosen for the additional layer-by-layer growth of new Fe-Ni alloy layers. One of the most important features of SLM-AM is the complex temperature history generated by the laser irradiation. Predicting the temperature history forms the foundation for the subsequent simulation of residual stress, microstructure, and magnetic property. Using the commercial FEA code ABAQUS \cite{version20136}, here we design and implement a non-linear transient thermal 3D model to obtain the laser induced global temperature history.

The governing equation for the energy balance of heat transfer in the SLM process is given as
\begin{equation}
k(\phi,T)T_{,ii} = \rho(\phi,T)C_\text{p}(\phi,T)\frac{\text{d}T}{\text{d}t},
\label{eq1}
\end{equation}
in which $T$ is the temperature, $\rho$ is the material density, $C_\text{p}$ is the specific heat capacity, $t$ is the time, and $k$ is the thermal conductivity. The initial condition for Eq. (\ref{eq1}) is
\begin{equation}
T(x_i,t_0)=T_0,
\label{eq2}
\end{equation}
in which $T_0$ is the ambient temperature 300 K. The temperature of the substrate bottom surface is set as constant $T_0=300$ K. On other surfaces, the thermal flux includes convection part $q_\text{con}$ and radiation part $q_\text{rad}$ which can be given as
\begin{equation}
q_\text{con}=h_c(T)(T-T_0)
\label{eq3}
\end{equation}
and
\begin{equation}
q_\text{rad}=\sigma_\text{sb}\epsilon_\text{sb}(T^4-T_0^4),
\label{eq4}
\end{equation}
respectively. In Eqs. (\ref{eq3}) and (\ref{eq4}), $h_\text{c}$ is the temperature dependent convective heat transfer coefficient, $T$ is the temperature of the corresponding surface, $\sigma_\text{sb}$ is the Stefan$-$Boltzmann constant, and $\epsilon_\text{sb}$ is the surface emissivity.

In Eq. \ref{eq1}, $\phi$ is a field variable to indicate the material state, i.e., whether the material has ever gone beyond the liquidus temperature $T_l$.
Each element stores its temperature $T$ and $\phi$. We set $\phi=0$ for a powder state and $\phi=1$ for a bulk state. $\phi$ is designed to realize the irreversible melting process from powder to bulk state by the subroutine USDFLD of ABAQUS. The powder elements are initialized with $\phi=0$. $\phi$ is changed from 0 to 1 upon melting and will retain 1 afterwards, i.e. the fused material can never go back to powder. $\phi$ of the substrate elements is initialized and always remains as 1.

The material parameters $\rho$ and $C_\text{p}$ are determined by the CALPHAD approach which is capable of predicting thermodynamically consistent properties. In the CALPHAD model, the Gibbs free energy per gram of one phase in a multicomponent system can be expressed as
\begin{equation}
G_\text{1g}(P,T)=\sum_i c_i G_i^0 + RT\sum_i c_i  \text{ln}c_i + G^\text{excess},
\label{eqgib}
\end{equation}
in which $c_i$ is the composition of element $i$ in the multicomponent system, $G_i^0$ the Gibbs free energy of pure element $i$, $R$ the gas constant, and $G^\text{excess}$ the excess Gibbs energy of mixing. Once the Gibbs energy is obtained from CALPHAD data, all other thermodynamic properties can be derived. For example, $\rho$ and $C_\text{p}$ can be calculated as
\begin{equation}
\rho(T)=\frac{1 \, \text{g}}{V_{1g}(P,T)}=\frac{1 \, \text{g}}{(\partial G_\text{1g}/\partial P)_T}
\label{eqrho}
\end{equation}
and
\begin{equation}
C_\text{p}(T)=-T (\partial G^2_{1g}/\partial T^2)_P,
\label{eqrho}
\end{equation}
as shown in Fig. \ref{f2}(b) and (c). The enthalpy per gram can be derived as
\begin{equation}
H_\text{1g}(T)=G_\text{1g} - T (\partial G_\text{1g}/\partial T)_P.
\label{eqh}
\end{equation}
The latent heat $L$ due to the change in enthalpy $\Delta H_\text{1g}$ of the system during the entire solid-liquid phase change can be calculated from Eq. (\ref{eqh}) as 291.14 J/g. It should be mentioned that according to the phase diagram of Fe-Ni alloy \cite{cacciamani2006critical}, the solidus temperature $T_s$ of permalloy with the composition around Fe$_{20}$Ni$_{80}$ is about 1,709 K, which is only 0.2 K lower than $T_l$. Therefore, a constant $L$ is taken here. For the system with a wide temperature region of solid-liquid mixture, $\Delta H_\text{1g}$ is a function temperature and the latent heat effects are often included in the temperature dependent effective specific heat \cite{smith2016thermodynamically,keller2017application}.

The thermal conductivity of powder ($k_\text{p}$) is usually very small and here is assumed to be $1\%$ of that of bull material ($k_\text{b}$) and slightly increase with temperature before melting. The powder density is assumed to be half of the bulk density. The powder specific heat is set the same as the bulk one. The temperature dependent material parameters used for thermal analysis are shown in Fig. \ref{f2}(a)--(c).

The interaction between the top surface and laser beam is simulated by a moving surface heat flux with a Gaussian distribution, i.e.
\begin{equation}
q_\text{a}=\frac{2\eta P_\text{a}}{\pi R_\text{a}^2} \text{exp}\left[-2\frac{\parallel \mathbf{r}-\mathbf{r}_0(v_\text{a},t) \parallel^2}{R_\text{a}^2} \right],
\label{eq5}
\end{equation}
where $\eta$ is the powder bed absorption coefficient with an assumed value of 0.5, $P_\text{a}$ is the laser power, $R_\text{a}$ is the laser beam radius, $\mathbf{r}$ is the coordinate of the point in the material, and $\mathbf{r}_0$ is a function of laser moving speed, and $v_\text{a}$ is the coordinate of laser beam center. The moving laser heat flux is dependent on the scanning strategy and can be realized by the subroutine DFLUX of ABAQUS. If not specified in the following, the laser beam parameters are chosen as $P_\text{a}=100$ W, $R_\text{a}=50$ $\mu$m, and $v_\text{a}=0.4$ m/s, according to the experimental work \cite{zhang2013magnetic}. This laser parameter may be difficult to be realized in industrial applications. Here we limit ourselves to the feasibility of the proposed computational scheme and will not focus on its application to the real industrial AM at the current stage.

\begin{figure}[!t]
\centering
\includegraphics[width=8.4cm]{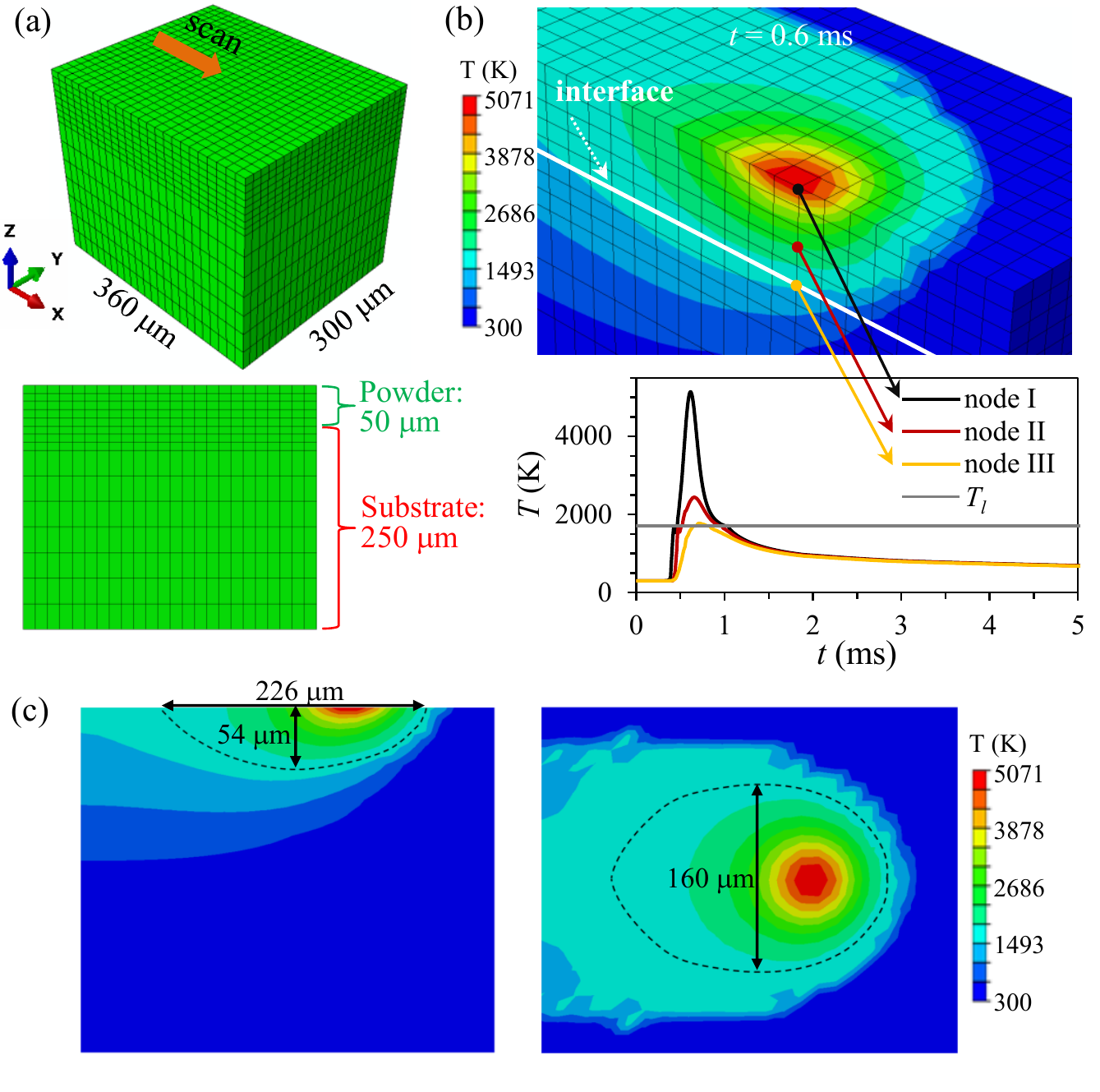}
\caption{Thermal analysis results for the single-track SLM scan along the middle line perpendicular to $y$ axis. (a) Model geometry and FE mesh, with a 50 $\mu$m thick powder layer. (b) Temporal evolution of temperature in three FE nodes.
(c) Temperature distribution on the cross-section of the molten pool at $t=0.6$ ms.}
\label{fT1}
\end{figure}

\begin{figure*}[!t]
\centering
\includegraphics[width=11cm]{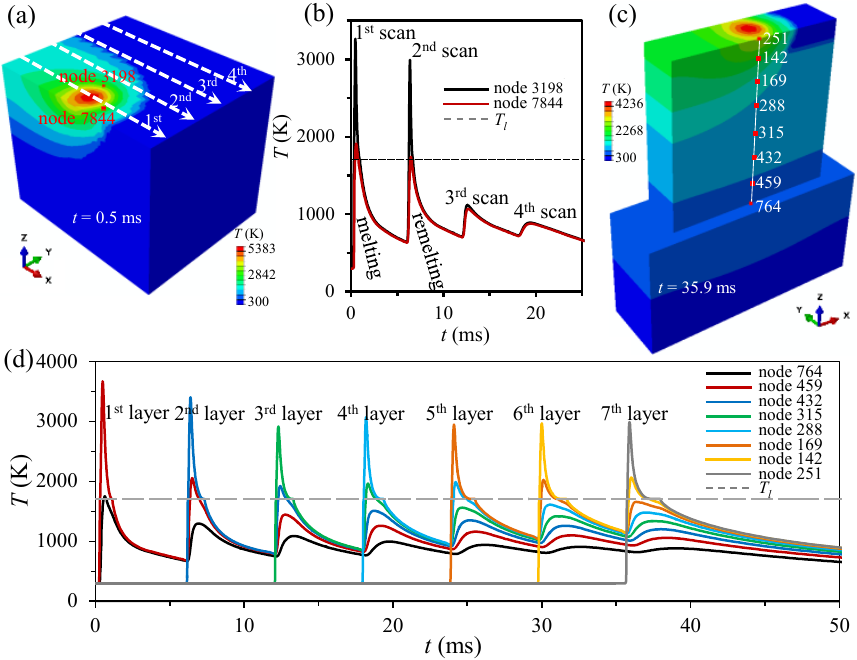}
\caption{Thermal analysis results for the multi-track SLM scan. In-plane four-track scan along $x$ axis: (a) temperature distribution at $t=0.5$ ms and (b) temporal evolution of temperature in the two FE nodes labeled in (a). Out-of-plane layer-by-layer scan: (c) temperature distribution at $t=35.9$ ms when the seventh layer is being built (d) temporal evolution of temperature in the FE nodes on the surface of each layer.}
\label{fT2}
\end{figure*}

Fig. \ref{fT1} shows the thermal history results for the single-track scan along the middle line perpendicular to $y$ axis. The powder layer is 50 $\mu$m thick and the substrate/powder model is discretized by hexahedral FE meshes, as displayed in Fig. \ref{fT1}(a).
Fig. \ref{fT1}(b) presents the temperature profile around the laser center at $t=0.6$ ms, as well as the temporal evolution of temperature at three FE nodes, e.g. one on the powder surface, one in the powder interior, and one in the substrate/deposit interface. It can be found that the powder at all these three nodes is quickly heated up to temperature above the melting point and then gradually cools down to the room temperature. The melting of node III in Fig. \ref{fT1}(b) ensures the good connection between the substrate and the deposited layer. By measuring the slope of the line connecting the maximum temperature of the peak to the temperature at 5 ms, an average cooling rate in the order of $10^5$ K/s can be obtained. Analogously, the average heating rate can be estimated to be in the order of $10^6$ K/s. These estimated rates indicate the feature of fast heating and cooling during SLM.

By examining the temperature profile at a certain time (e.g. $t=$0.6 ms), the molten pool geometry can be obtained, as shown in Fig. \ref{fT1}(c). The length and depth of the molten pool is estimated as 226 and 54 $\mu$m, respectively. The molten pool appears like a comet tail, whose asymmetry could be attributed to the laser movement, as well as the temperature and material state dependent thermal conductivity. In front of the laser, the material is in powder state with low thermal conductivity, thus leading to slow heat transfer and high temperature gradient. On the contrary, in rear of laser the bulk material state possesses higher thermal conductivity and wide temperature distribution. A similar melt pool geometry is also reported in literature on nonmagnetic materials \cite{keller2017application}.

In contrast to the single-track scan for a strip-like material, the in-plane and out-of-plane multi-track scans are simulated to build a one-layer and multi-layer bulk material, respectively. The associated thermal results are presented in Fig. \ref{fT2}. 
During the multi-track scanning process, an idle time of 5 ms between the completion of one track and the beginning of the subsequent track is assumed. The idle time is demonstrated to be important \cite{costa2005rapid}, but its optimization is out of the scope here. 
Cyclic heating and cooling is remarkable during the multi-track scanning process, as shown in Fig. \ref{fT2}(b) and (d). The temporal evolution of temperature at the interfacial nodes between the first and second scanning track (nodes 3198 and 7844 marked in Fig.\ref{fT2}(a)) indicates four heating-cooling cycles. Especially, these two material nodes experience notable melting in the first scan and remelting in the second scan. The remelting means that during the second scan the molten pool can extend to the previously deposited track, resulting in good inter-track bonding. During the third and fourth scans, although these two nodes do not melt again, heating and cooling with a temperature change around 500 K still occurs and may raise debonding and thermal fatigue issues.
For the out-of-plane layer-by-layer multi-track scan, the continuous addition of powder is considered by using the element deactivation and activation, i.e. successive discrete addition of new elements into the new scanning track at the beginning of each time step. By using the heat accumulation effect in SLM process \cite{li2014parametric} in which the heat stored in the previous layer affects the next processing layer and induces overheating, the laser power can be varied layer by layer, i.e. large powder for the initial layers and small power for the subsequent layers. Fig. \ref{fT2}(c) and (d) presents the typical thermal results in the case of 100 W for the first layer, 60 W for the second layer, and 50 W for the other five layers. The temperature evolution of nodes (marked in Fig. \ref{fT2}(c)) at the surface of each layer in Fig. \ref{fT2}(d) shows melting and remelting process, indicating the possibility of inter-layer bonding and the integration of deposited layers into a bulk material.
\section{Mechanical analysis}
The mechanical analysis is subsequently performed independently, since it is reasonable that the mechanical response has a neglectable effect on the thermal history, and the thermal and mechanical analyses are weakly coupled. The analysis is based on the thermal history dependent quasi-static mechanical model, which takes the above thermal results as thermal loads. The governing equation for the stress equilibrium is
\begin{equation}
\sigma_{ij,j} = 0,
\label{eq6}
\end{equation}
in which $\sigma_{ij}$ is the stress. For the mechanical boundary condition, the rigid body motion is restricted and the substrate bottom surface is free to deform. The mechanical constitutive law can be given as
\begin{equation}
\sigma_{ij} = C_{ijkl}(\phi,T)\epsilon_{kl}^\text{e},
\label{eq7}
\end{equation}
where the elastic tensor $C_{ijkl}$ can be expressed by temperature dependent elastic modulus (Fig. \ref{f2}(e)) and a Poisson ratio of 0.33 for the isotropic material behavior considered here. The total strain is decomposed into elastic strain $\epsilon_{ij}^\text{e}$, plastic strain $\epsilon_{ij}^\text{p}$, and thermal strain $\epsilon_{ij}^\text{T}$, i.e.
\begin{equation}
\epsilon_{ij} = \epsilon_{kl}^\text{e} + \epsilon_{kl}^\text{p}+ \epsilon_{kl}^\text{T}.
\label{eq8}
\end{equation}
Thermal strain is given by $\epsilon_{ij}^\text{T}=\alpha(\phi,T)(T-T_0)\delta_{ij}$ in which $\alpha$ is the thermal expansion coefficient, $T_0$ is the initial temperature, and $\delta_{ij}$ is the Kronecker delta.
$\alpha$ is calculated by the CALPHAD approach through the Gibbs free energy in Eq. (\ref{eqgib}), i.e.
\begin{equation}
\alpha(T)=\frac{1}{3}\frac{1}{V}\frac{\partial V}{\partial T} = \frac{1}{3}\frac{1}{(\partial G_\text{1g}/\partial P)_T} \frac{\partial^2G_\text{1g}}{\partial P \partial T},
\label{eqalpha}
\end{equation} 
as shown in Fig. \ref{f2}(d). For the computation of plastic strain, the linear isotropic hardening model and von Mises yield criterion are used. The yield function is computed as
\begin{equation}
f(\sigma_{ij},\sigma_\text{Y}^\text{0},\sigma_\text{Y}^\text{h}) = \sigma_\text{mises} - (\sigma_\text{Y}^0 + \sigma_\text{Y}^\text{h}).
\label{eq9}
\end{equation}
In Eq. \ref{eq9}, $\sigma_\text{mises}$ is the von Mises stress calculated from the stress tensor $\sigma_{ij}$. $\sigma_\text{Y}^0(\phi,T)$ is the initial yield stress without equivalent plastic strain, as shown by the left curve in Fig. \ref{f2}(f). $\sigma_\text{Y}^\text{h}$ represents the hardening and linearly correlates with the equivalent plastic strain $\epsilon_\text{e}^\text{p}$ through the hardening coefficient $E^\text{h}(\phi,T)$ (right curve in  Fig. \ref{f2}(f)), i.e. $\sigma_\text{Y}^\text{h}=E^\text{h} \epsilon_\text{e}^\text{p}$. The plastic strain is computed by combing the yield criterion in Eq. \ref{eq9} and the Prandtl--Reuss flow rule.

\begin{figure}[!b]
\centering
\includegraphics[width=8.4cm]{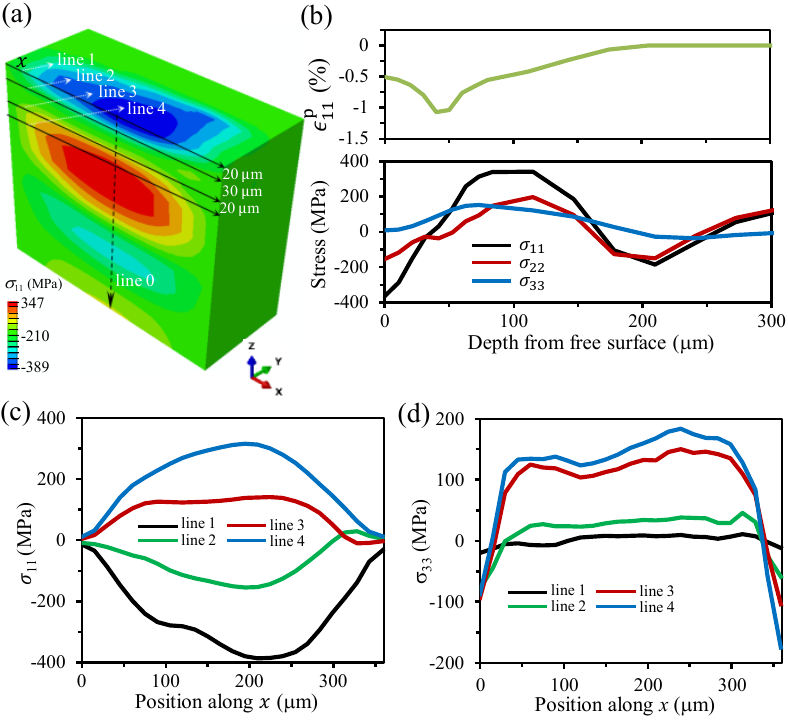}
\caption{Calculated residual stress when the single-track laser beam has been switched off and the temperature has equilibrated to 300 K. (a) Contour (a vertical $y$-midplane cutting through the model) of stress $\sigma_{11}$ distribution. (b) Plastic strain $\epsilon^\text{p}_{11}$ and stress component distribution along line 0 displayed in (a). Distribution of stress (c) $\sigma_{11}$ and (d) $\sigma_{33}$ along the four lines marked in (a).}
\label{fS1}
\end{figure}

\begin{figure}[!b]
\centering
\includegraphics[width=8.4cm]{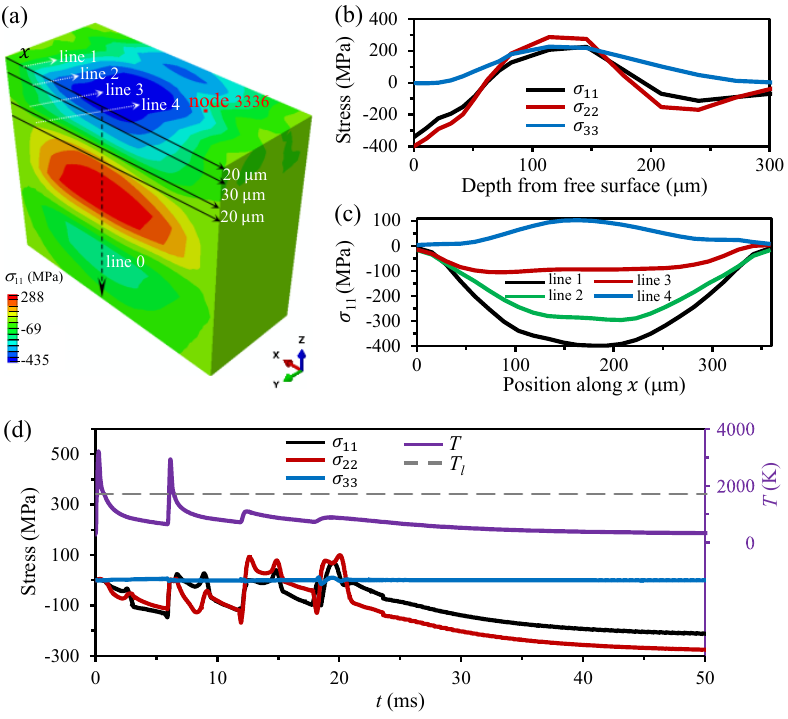}
\caption{Calculated residual stress when the in-plane multi-track laser beam has been switched off and the temperature has equilibrated 300 K. (a) Contour (a vertical $y$-midplane cutting through the model) of stress $\sigma_{11}$ distribution. (b) Stress component distribution along line 0 marked in (a). (c) Distribution of stress $\sigma_{11}$ along the four lines marked in (a). (d) Temporal evolution of stress and temperature at surface node 3336 marked in (a).}
\label{fS2}
\end{figure}

Fig. \ref{fS1} gives the results of residual stress in the case of single-track scan. The distribution of residual stress $\sigma_{11}$ along the scanning direction is shown in Fig. \ref{fS1}(a). It is apparent from the contour plot that compressive $\sigma_{11}$ appears in the deposited layer and tensile $\sigma_{11}$ in the substrate region close to the deposited layer. Fig. \ref{fS1}(b) shows the distribution of through-thickness residual stresses and plastic strain $\epsilon^\text{p}_{11}$ (along the line 0 marked in Fig. \ref{fS1}(a)). It can be seen that both the stress components $\sigma_{11}$ and $\sigma_{22}$ are compressive in the deposited layer, but change from tensile to compressive in the substrate. The stress component $\sigma_{33}$ through the thickness is relatively small.
Furthermore, the residual stress distribution along $x$ direction in the midplane is examined in terms of the 4 lines defined in Fig. \ref{fS1}(a).
It can be seen from Fig. \ref{fS1}(c) that $\sigma_{11}$ gradually changes from compressive along line 1 to tensile along line 4.
The compressive $\sigma_{11}$ on the free surface of the deposit are caused by the steep temperature gradient, i.e. the expansion of the hotter top-layer material is prohibited by the underlying material with much lower temperature. In addition, the thermally induced plastic strain should be responsible for the residual stress; because pure elasticity with homogeneous material parameters under no external constraint will not generate residual stress after cooling down to a uniform temperature. The distribution of plastic strain $\epsilon^\text{p}_{11}$ in the deposited layer, as shown in Fig. \ref{fS1}(b), also favors the compressive $\sigma_{11}$ after cooling down to the room temperature. The tensile stresses in the substrate/deposit interface can be attributed to the cooling down of the molten material \cite{mercelis2006residual} and the self-balance of the whole structure. Generally, compressive residual stresses in the top part of the deposit are favorable for increasing the load resistance and preventing crack growth. But tensile residual stresses in the bottom part of the deposit are disadvantageous since they could reduce the load resistance and accelerate crack growth.

\begin{figure}[!b]
\centering
\includegraphics[width=8.4cm]{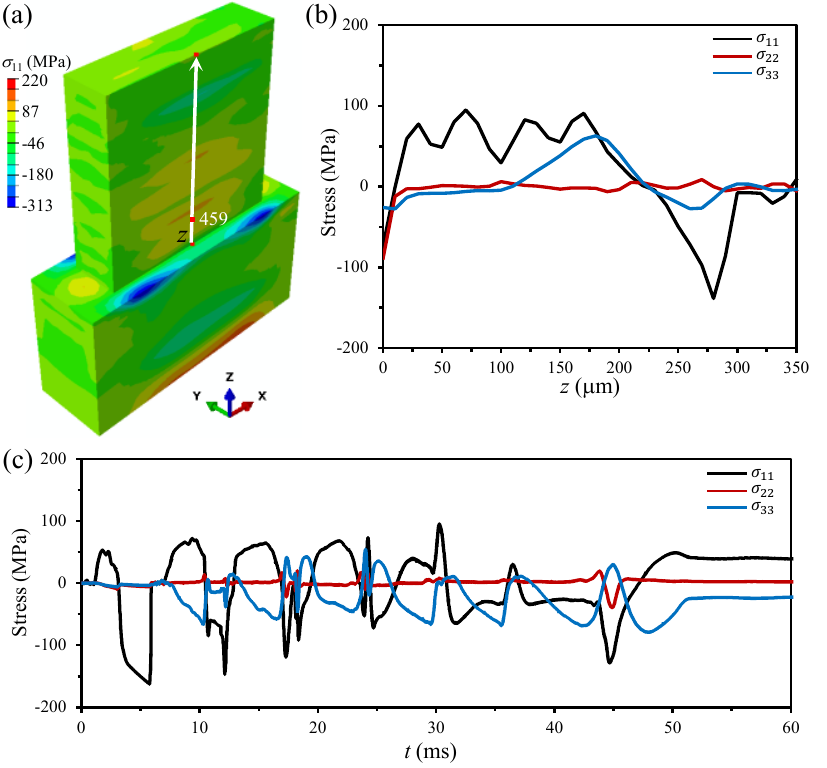}
\caption{Calculated residual stress when the out-of-plane multi-track laser beam has been switched off and the temperature has equilibrated 300 K. (a) Contour of stress $\sigma_{11}$ distribution. (b) Stress component distribution along the line marked in (a). (c) Temporal evolution of stress at an inter-layer node 459 marked in (a).}
\label{fS3}
\end{figure}

The residual stress distribution in the multi-track scan is in Figs. \ref{fS2} and \ref{fS3}. For the in-plane four-track scan in Fig. \ref{fS2}, the stress distribution in Fig. \ref{fS2}(b) and (c) is similar to that in the single-track case in Fig. \ref{fS1}. But the tensile stress $\sigma_{11}$ in the substrate is lower.
For the out-of-plane layer-by-layer multi-track scan in Fig. \ref{fS3}, the residual stress is even much lower. The through-thickness stress distribution in Fig. \ref{fS3}(c) shows a average residual stress around 50 MPa, much smaller than that in the single-track and in-plane multi-track scan. The reason could be related to the partial relief of stress under reheating and cooling during by the subsequent laser scanning for depositing the adjacent layers.
The cyclic heating and cooling during the multi-track scan also result in cyclic stress history, as depicted by Fig. \ref{fS2}(d) and  Fig. \ref{fS3}(c). It can be seen that all stress components are almost zero when the liquidus temperature is reached. Most importantly, $\sigma_{11}$ and $\sigma_{22}$ evolution in Fig. \ref{fS2}(d) and  Fig. \ref{fS3}(c) manifest that the marked nodes experience somewhat cyclic tension and compression along $x$ and $y$ directions. Since the marked nodes are in the interface of adjacent layers, the cyclic tension and compression could weaken the interface bonding, or even lead to interface failure.

It should be mentioned that in the above thermal and mechanical analysis, the calculation methodology for the magnetic FeNi material is similar to that for the conventional alloys. No special treatment is proposed to deal with the magnetic contribution to the temperature and stress/strain. This is an approximation which is reasonable due to the following two aspects. Firstly, the influence of magnetic properties of FeNi on the heat-transfer thermal analysis is negligible. Secondly, the magnetostrictive coefficient of FeNi is in the order of 10$^{-7}$ to 10$^{-6}$, which is so small that the effect of magnetization on the stress/strain can be neglected when compared to the effect of thermal expansion \cite{shu2004micromagnetic}. So the mechanical analysis can be performed by using the similar method for conventional alloys. However, if one deals with giant magnetostrictive materials (e.g. Terfenol-D with 10$^{-3}$ order of magnitude of magnetostriction), the magnetization contribution in the stress/strain calculation cannot be ignored \cite{shu2004micromagnetic}.

\section{Microstructure evolution}
Microstructure plays a critical role in the property of the products processed by SLM-AM and is required to be predicted if possible. Here we attempt to predict the microstructure evolution in Fe-Ni permalloy during SLM process by using the thermal history and the temperature-dependent phase fraction estimated from CALPHAD. The CALPHAD method is capable of predicting not only thermodynamical properties for material design, but also microstructural evolution through comprehensive physical models of materials processing \cite{lukas2007computational}. For example, recently CALPHAD has been combined with phase-field simulation \cite{keller2017application,perron2017matching} and heat-transfer simulation \cite{smith2016linking,smith2016thermodynamically} to predict the process-phase relationships.

\begin{figure}[!b]
\centering
\includegraphics[width=8cm]{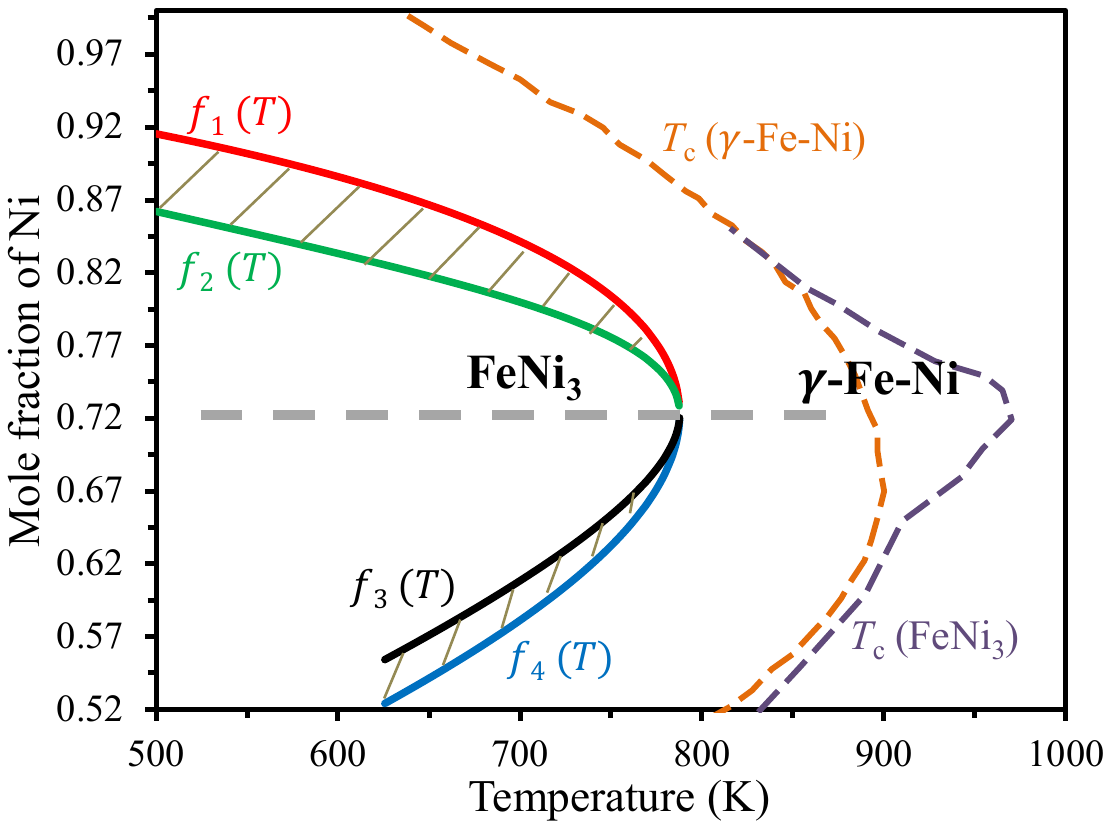}
\caption{CALPHAD informed phase boundary curves around a Ni atomic percent of 80$\%$. The shadow region indicates the coexistence of $\gamma$-Fe-Ni and FeNi$_3$ phases.
 The Curie temperature cures for $\gamma$-Fe-Ni and FeNi$_3$ are also presented \cite{swartzendruber1991fe}. }
\label{f3}
\end{figure}

\begin{figure*}[!t]
\centering
\includegraphics[width=12cm]{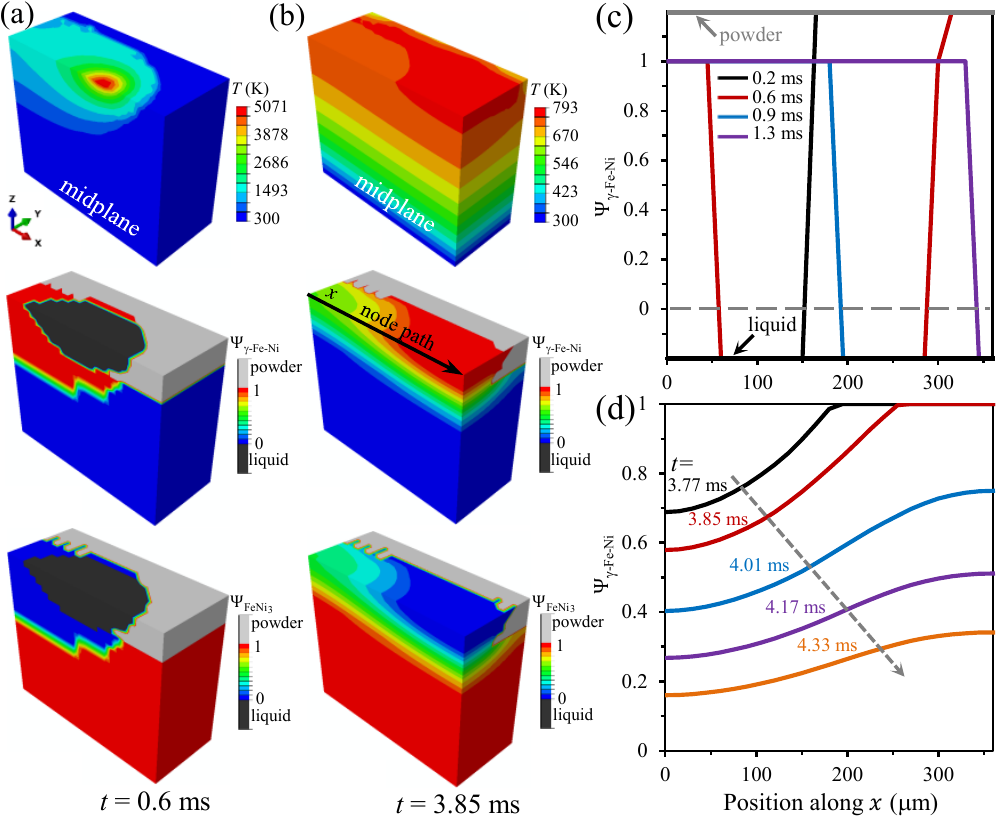}
\caption{Phase evolution during the single-track SLM scan, with the powder composition $x_\text{Ni}=0.8$. Snapshots (a vertical $y$-midplane cutting through the model) of the predicted distribution of temperature, liquid, powder, $\gamma$-Fe-Ni phase fraction ${\rm \Psi}_{\gamma\text{-Fe-Ni}}$, and FeNi$_3$ phase fraction ${\rm \Psi}_{\text{FeNi}_3}$ at time (a) 0.6 ms and (b) 3.85 ms. (c) and (d) ${\rm \Psi}_{\gamma\text{-Fe-Ni}}$ distribution along the node path marked in (b).}
\label{fp1}
\end{figure*}

Based on the thermodynamic data of Fe-Ni alloy in the CALPHAD software Thermo-Calc \cite{andersson2002thermo}, we can predict the phase at any temperature for a given Fe-Ni composition. Since in present work we are interested in the magnetic Fe-Ni permalloy whose composition is around Fe$_{20}$Ni$_{80}$, only the Thermo-Calc calculated results around Fe$_{20}$Ni$_{80}$ are shown in Fig. \ref{f3}. The phase distribution of permalloy region in Fig. \ref{f3} is partitioned into four parts by four curves $f_1 (T)$, $f_2 (T)$, $f_3 (T)$ and $f_4 (T)$, which can be either fitted by piecewise-smooth functions or directly used as scattered data from Thermo-Calc output. For numerical study here, we extract scatter data from these four curves and interpolate $f_i(T)$ values at any temperature. The peak temperature is around 787 K and the corresponding mole fraction of Ni ($x_\text{Ni}$) is around 0.72.

As a first attempt and for simplicity, here we only consider FeNi$_3$ and $\gamma$-Fe-Ni phases. The real experimental case should be more complicated, whose comprehensive modeling cannot be achieved within one step and will be continuously explored in the near future. In this way, we can find from Fig. \ref{f3} that $\gamma$-Fe-Ni phase exists at temperature between 761 K and 1709 K for $x_\text{Ni}=0.8$. Above 1709 K, only liquid exists. In the region bounded by curves $f_2(T)$ and $f_3(T)$, only FeNi$_3$ exists. When $x_\text{Ni}$ is 0.72, single phase appears at any temperature and the stoichiometric is the same as the powder. In the shadow region bounded by curves $f_1(T)$ and $f_2(T)$, and curves $f_3(T)$ and $f_4(T)$, the coexistence of FeNi$_3$ and $\gamma$-Fe-Ni phases occurs. For the phase-coexistence region with $0.72<x_\text{Ni}<0.90$, by using the lever rule, we can calculate the FeNi$_3$ phase mole fraction ${\rm \Psi}_{\text{FeNi}_3}$ and the $\gamma$-Fe-Ni phase mole fraction ${\rm \Psi}_{\gamma\text{-Fe-Ni}}$ by applying functions $f_1(T)$ and $f_2(T)$. Similarly, for the case $0.52<x_\text{Ni}<0.72$, ${\rm \Psi}_{\text{FeNi}_3}$ and ${\rm \Psi}_{\gamma\text{-Fe-Ni}}$ can be calculated from functions $f_3(T)$ and $f_4(T)$. Then if $x_\text{Ni}$ of the initial powder is given, at any time $t$ the level rule reads

\begin{equation} 
 {\rm \Psi}_{\text{FeNi}_3}[T(t)] =  \left\{
\begin{array}{rcl}
  \displaystyle{\frac{f_1[T(t)]-x_\text{Ni}}{f_1[T(t)]-f_2[T(t)]}}, \,\, {0.72<x_\text{Ni}<0.90;} \\[1.1em]
  \displaystyle{\frac{x_\text{Ni}-f_4[T(t)]}{f_3(T)-f_4[T(t)]}}, \,\, {0.52<x_\text{Ni}<0.72;}
\end{array}  \right.
\label{eq10}
\end{equation}
and
\begin{equation}
{\rm \Psi}_{\gamma\text{-Fe-Ni}}[T(t)] =  \left\{
\begin{array}{rcl}
 \displaystyle{\frac{x_\text{Ni}-f_2[T(t)]}{f_1[T(t)]-f_2[T(t)]}}, \,\, 0.72<x_\text{Ni}<0.90; \\[1.1em]
 \displaystyle{\frac{f_3[T(t)]-x_\text{Ni}}{f_3[T(t)]-f_4[T(t)]}}, \,\, 0.52<x_\text{Ni}<0.72.
\end{array}  \right.
\label{eq11}
\end{equation}
 
By using Fig. \ref{f3}, Eqs. \ref{eq10} and \ref{eq11}, and the temperature evolution available at each FE node from the thermal analysis, the concurrent and spatially varying microstructure evolution could be predicted. Taking the initial powder with composition $x_\text{Ni}=0.8$ as an example, Fig. \ref{fp1} shows the temporal evolution of powder, liquid phase, and the phase fraction of $\gamma$-Fe-Ni and FeNi$_3$ in the case of single-track scan. At $t=0.6$ ms (Fig. \ref{fp1}(a)), the laser beam is on and the temperature gradient is very large. In front of liquid phase (black color), powder is still there and no phase forms. Due to the high temperature around the liquid phase, the already deposited layer behind the liquid phase and the substrate region close to the liquid phase possess only $\gamma$-Fe-Ni phase.
Because of the neglectable difference between the liquidus and solidus temperatures, it can be seen from Fig. \ref{fp1}(a) that the region with liquid-solid coexistence is almost unobservable and the interface between liquid and solid is sharp. This result is different from the previous work on stainless steel 316L with a solidus-liquidus temperature difference around 120 K calculated from CALPHAD \cite{smith2016thermodynamically}. Meanwhile, the region with coexistence of $\gamma$-Fe-Ni and FeNi$_3$ phases is also very narrow and only exists in substrate. When the time goes to 0.9 ms, the laser beam is off. As presented in Fig. \ref{fp1}(b) at 3.85 ms, the temperature is lower than 793 K, no liquid phase remains, and the deposited layer completely covers $x$ direction. In the deposited layer, there are wide regions where $\gamma$-Fe-Ni and FeNi$_3$ phases coexist. More specifically, Fig. \ref{fp1}(c) and (d) depicts the phase fraction of $\gamma$-Fe-Ni (${\rm \Psi}_{\gamma\text{-Fe-Ni}}$) along the surface node path marked in Fig. \ref{fp1}(b) at various times. It can be found from Fig. \ref{fp1}(c) that within 1.3 ms, $\gamma$-Fe-Ni (${\rm \Psi}_{\gamma\text{-Fe-Ni}}=1$), liquid, and powder exist along the node path, but the interface between them is extremely narrow. During the cooling down process in Fig. \ref{fp1}(d), ${\rm \Psi}_{\gamma\text{-Fe-Ni}}$ is found to be between 0 and 1 in a wide region along the node path, indicating the obvious coexistence of $\gamma$-Fe-Ni and FeNi$_3$ phases in the deposited layer.
In order to clearly show the mixture of FeNi$_3$ and $\gamma$-Fe-Ni phases, the time in Fig. \ref{fp1}(d) is selected so that the corresponding temperature falls into the region of phase coexistence.
Fig. \ref{fp1} also indicate that after cooling down to 300 K, the final state is the ordered intermetallic compound FeNi$_3$ phase. For the composition Fe$_{20}$Ni$_{80}$, the additional Ni element is present with the formation of solid solution in FeNi$_3$ phase. The conclusion with final FeNi$_3$ phase also agrees with the experimental X-ray diffraction results from the literature \cite{zhang2013magnetic}, in which it is demonstrated that FeNi$_3$ phase is readily identified and the impurity like other Fe-Ni intermetallic compounds or Fe-Ni simple substance cannot be found in samples.

\begin{figure*}[!t]
\centering
\includegraphics[width=12cm]{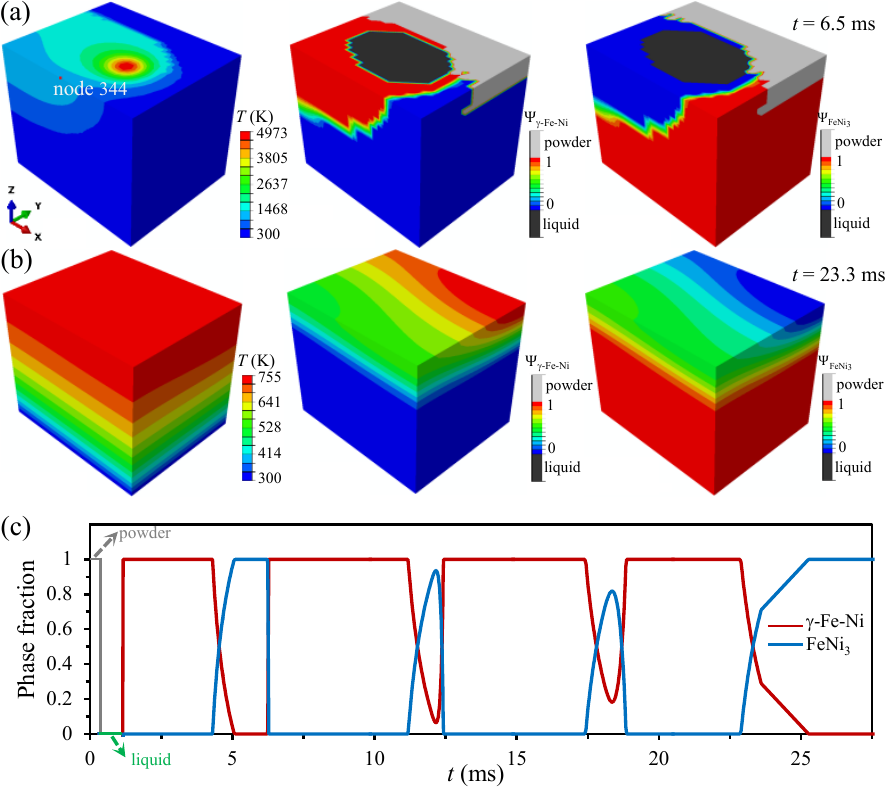}
\caption{Phase evolution during the in-plane multi-track SLM scan along $x$ axis, with the powder composition $x_\text{Ni}=0.8$. Snapshots of the predicted distribution of temperature, liquid, powder, $\gamma$-Fe-Ni phase fraction ${\rm \Psi}_{\gamma\text{-Fe-Ni}}$, and FeNi$_3$ phase fraction ${\rm \Psi}_{\text{Fe-Ni}_3}$ at time (a) 6.5 ms and (b) 23.3 ms. (d) Temporal evolution of phase fraction of liquid, $\gamma$-Fe-Ni, and FeNi$_3$ at the node 344 marked in (a).}
\label{fp2}
\end{figure*}

\begin{figure*}[!t]
\centering
\includegraphics[width=12cm]{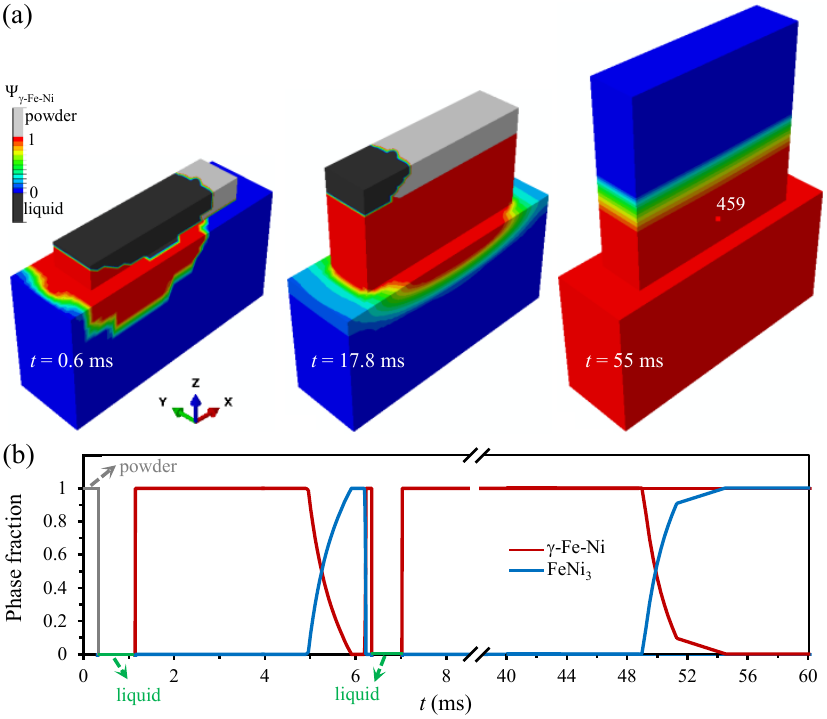}
\caption{Phase evolution during the out-of-plane multi-track SLM scan, with the powder composition $x_\text{Ni}=0.8$. (a) Snapshots of the predicted $\gamma$-Fe-Ni phase fraction ${\rm \Psi}_{\gamma\text{-Fe-Ni}}$ at different times. (d) Temporal evolution of phase fraction of liquid, $\gamma$-Fe-Ni, and FeNi$_3$ at the node 459 marked in (a).}
\label{fp3}
\end{figure*}

Fig. \ref{fp2} shows the phase evolution during the in-plane multi-track scan. At $t=6.5$ ms in Fig. \ref{fp2}(a), the laser beam is around the center of the second track and the powder state remains in front of the melt pool. A narrow interface between $\gamma$-Fe-Ni and FeNi$_3$ is found in both the substrate and the first-track deposit.  During the cooling down of the whole sample at $t=23.3$ ms in Fig. \ref{fp2}(b), $\gamma$-Fe-Ni/FeNi$_3$ interface is significantly broadened and a large region in the deposited layer contains two phases. It is obvious that in the contour plots in Fig. \ref{fp2}(a) and (b), the subsequent laser scan induces phase changes in the previously deposited layer. More specifically, Fig. \ref{fp2}(c) depicts the temporal evolution of phase fraction at the edge node marked in Fig. \ref{fp2}(a). It can be seen that this material node changes from powder to liquid during the first-track scan, and then experiences cyclic phase changes between $\gamma$-Fe-Ni and FeNi$_3$ during the subsequent scans. The cyclic phase change is originated from the cyclic temperature history as discussed in Fig. \ref{fT2}(b).

Similarly, Fig. \ref{fp3} presents the phase evolution during the out-of-plane multi-track scan. As shown in Fig. \ref{fp3}(a), during the deposition of the first layer at $t=0.6$ ms, liquid, powder, $\gamma$-Fe-Ni and FeNi$_3$ all exist and the interfaces among them are very sharp. At the beginning of depositing the third layer ($t=17.8$ ms), the previously deposited two layers are $\gamma$-Fe-Ni and a wide region with the coexistence of $\gamma$-Fe-Ni and FeNi$_3$ appears in the substrate. During the cooling process ($t=55$ ms) after the completion of seven-layer deposition, a layered region possessing both $\gamma$-Fe-Ni and FeNi$_3$ emerges in the built vertical structure. The temporal evolution of phase fraction at the edge node (as marked in Fig. \ref{fp3}(a)) in Fig. \ref{fp3}(b) shows that the material node experiences cyclic phase changes.

\begin{figure}[!b]
\centering
\includegraphics[width=8.4cm]{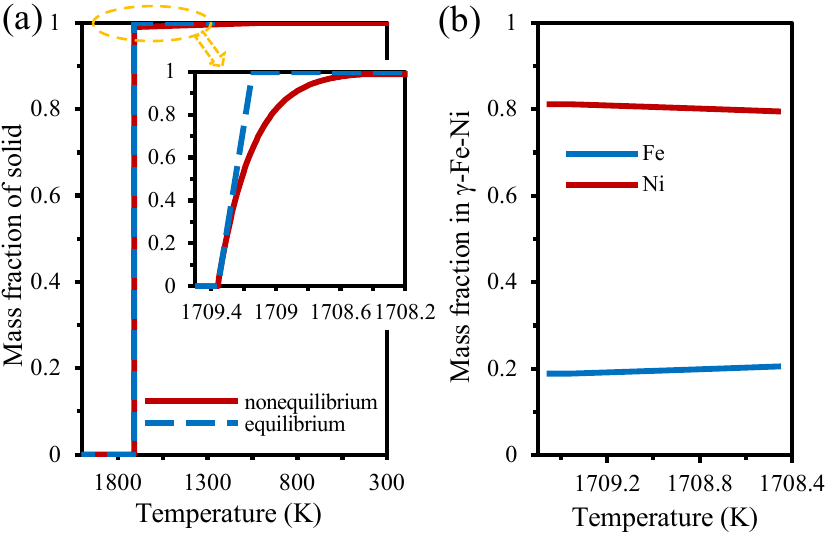}
\caption{(a) Fraction of solid as a function of temperature: equilibrium result from the phase diagram and nonequilibrium result from the Scheil-Gulliver solidification model through the CALPHAD approach. (b) Mass fraction of Fe and Ni in $\gamma$-Fe-Ni phase during Scheil-Gulliver solidification.}
\label{fnonequi}
\end{figure}

It should be noted that in the above analysis, the liquid-solid coexistence is ignored and the solidification be assumed to occur instantly due to the neglectable difference between liquidus and solidus temperatures from the equilibrium phase diagram. However, nonequilibrium solidification process could happen in the real case. In order to deal with the nonequilibrium solidification, we utilize the Scheil-Gulliver model \cite{gulliver1913quantitative,scheil1942remarks} which has been implemented in the Thermo-Calc package. The model assumes perfect mixing in the liquid and no diffusion in the solid phase. As introduced by Scheil in 1942 \cite{scheil1942remarks}, the partition coefficient ($\kappa$) during solidification can be defined as the ratio of the local composition of the solid phase $C_\text{S}$ to that of the liquid phase $C_\text{L}$, $C_\text{S}/C_\text{L}$ as determined from the phase diagram. If $C_\text{0}$ is the starting composition and $f_\text{S}$ is the fraction of solid, $C_\text{S}$ can be obtained through the following equation
\begin{equation}
(C_\text{L}-C_\text{S})\text{d}f_\text{S}=(1-f_\text{S})\text{d}C_\text{L}.
\label{scheil}
\end{equation}
The solution of Eq. (\ref{scheil}) is $C_\text{S}=\kappa C_0 (1-f_\text{S} )^{\kappa-1}$. The calculated fraction of solid ($f_\text{S}$) in both equilibrium and nonequilibrium solidification for FeNi powder with composition $x_\text{Ni}=0.8$ is shown in Fig. \ref{fnonequi}. It can be found from Fig. \ref{fnonequi}(a) that within a temperature range of 300--2000 K, the difference between the equilibrium result and the nonequilibrium result is hardly observable. Minor difference only appears around the liquidus and solidus temperatures. As shown in the inset of Fig. \ref{fnonequi}(a), the liquid-solid coexistence region is about 0.2 K in the equilibrium result, and is about 0.9 K in the nonequilibrium result. The nonequilibrium result gives more accurate prediction on the fraction of solid within the temperature range of 0.9 K (from 1709.36 to 1708.46 K in Fig. \ref{fnonequi}(a)). However, the deviation of the above equilibrium results from the nonequilibrium result only exists in this small temperature range of 0.9 K. In addition, Fig. \ref{fnonequi}(b) indicates that during the nonequilibrium solidification, microsegregation including Ni depletion and Fe enrichment in $\gamma$-Fe-Ni occurs, but it is extremely weak.

Solid phase transformation during the SLM processing induced heating and cooling is also important, i.e. the transformation between $\gamma$-Fe-Ni phase and FeNi$_3$ phase for the FeNi alloy with a composition $x_\text{Ni}$ around 0.8. It should be mentioned that the results on the transformation between $\gamma$-Fe-Ni and FeNi$_3$ at low temperatures in Fig. \ref{fp1}(d), Fig. \ref{fp2}(c), and Fig. \ref{fp3}(b) are obtained by using the phase diagram and lever principle without the consideration of nonequilibrium effect. This is an extremely rough approximation. More accurate or quantitative modeling and computation of heating and cooling induced phase transformations between $\gamma$-Fe-Ni and FeNi$_3$ phases with nonequilibrium effect are highly recommended, but are still challenging. On the one hand, in contrast to the disordered $\gamma$-Fe-Ni solid solution, FeNi$_3$ is a ordered phase whose modeling is difficult due to the complexity of involved sublattices and the lack of mobility parameters. On the other hand, phase-field modeling provides the feasibility of simulating microstructure evolution during solid phase transformation \cite{chen2002phase}, but it requires the comprehensive thermodynamic and kinetic data of both disordered $\gamma$-Fe-Ni and ordered FeNi$_3$ phases which are unfortunately not readily available at the current stage. Future efforts have to be made towards modeling microstructure evolution and nonequilibrium effect in heating and cooling induced phase transformations between $\gamma$-Fe-Ni and FeNi$_3$.
\section{Magnetic properties}
As a functional magnetic material, Fe-Ni permalloy is featured by its magnetic properties, in contrast to the structure materials  with the merit of strength and ductility. Therefore, the focus of property prediction here is different from that in SLM processed structural materials.
Generally speaking, the phase transformation during laser scanning process will influence the intrinsic magnetic properties of the formed phases, such as Curie temperature $T_\text{c}$, saturation magnetization $M_\text{s}$, magnetocrystalline anisotropy $K_\text{a}$, and magnetostriction $\lambda_\text{s}$. Since magnetic properties are concerned only when the temperature is below $T_\text{c}$ (above which a paramagnetic state exits), the Curie temperature curves are important and thus are plotted in the phase diagram, as shown in Fig. \ref{f3}. It can be found that FeNi$_3$ possesses higher $T_\text{c}$ than $\gamma$-Fe-Ni. It means that if $\gamma$-Fe-Ni is transformed into FeNi$_3$ in the phase coexistence region, $T_\text{c}$ of the system will be increased. But the quantitative calculation of $T_\text{c}$ in this phase coexistence region is still challenging. Taking the composition $x_\text{Ni}=0.85$ as an example, the formed phases will be magnetic only below 815 K. In addition, $M_\text{s}$ and $\lambda_\text{s}$ of both $\gamma$-Fe-Ni and FeNi$_3$ with the same composition $x_\text{Ni}$ are similar \cite{swartzendruber1991fe}. As typical soft magnetic materials, both $\gamma$-Fe-Ni and FeNi$_3$ have extremely small magnetocrystalline anisotropy and thus negligible $K_\text{a}$ values. Therefore, under the fixed composition $x_\text{Ni}$ around 0.8, the phase transformation below the Curie temperature curves in Fig. \ref{f3} will not significantly influence the intrinsic magnetic properties $M_\text{s}$, $K_\text{a}$, and $\lambda_\text{s}$.
In contrast, magnetic coercivity is an extrinsic magnetic property, which is a very important indicator for the assessment of magnetic materials and is the manifestation of these intrinsic magnetic properties combined with extrinsic factors such as shape, stress/strain, etc. To this end, we attempt to calculate the magnetic properties in the SLM processed Fe-Ni alloy by using micromagnetic simulations, with the consideration of SLM induced residual stress.

In the micromagnetic framework \cite{fidler2000micromagnetic,manfred2003micromagnetism,yi2016micromagnetic}, the spatial distribution of the magnetization is described as $\mathbf{m}M_\text{s}$ in which $\mathbf{m}$ is the unit vector of the magnetization direction and $M_\text{s}$ is the saturation magnetization. Micromagnetic model is in the framework of continuum theory that handles magnetization processes on a length scale that is small enough to resolve the transition of the magnetization within domain walls but large enough to replace the atomic magnetic moments by a continuous function of position. The free energy density of a magnetic body is defined as
\begin{equation}
E^\text{tot} = E^\text{exch} + E^\text{ani} + E^\text{demag} + E^\text{Zeeman} + E^\text{m-ela}.
\label{eqEtot}
\end{equation}
In Eq. (\ref{eqEtot}), the exchange energy
\begin{equation}
E^\text{exch}=A_\text{e}\|\nabla \mathbf{m}\|^2
\label{eqexch}
\end{equation}
is related to the gradient of $\mathbf{m}$ and contributes to the domain wall energy, with $A_\text{e}$ denotes the exchange parameter which is taken as 13 pJ/m for Fe-Ni Permalloy \cite{agramunt2014controlling}. The magnetocrystalline anisotropy energy
\begin{equation}
E^\text{ani} = \left \{
             \begin{array}{lr}
             K_\text{a}(m_1^2m_2^2+m_2^2m_3^2+m_3^2m_1^2), &  \text{cubic anisotropy} \\
             K_\text{a} [1-(\mathbf{m} \cdot \mathbf{u})^2 ], & \text{uniaxial anisotropy}
             \end{array}
\right.
\label{eqani}
\end{equation}
physically means that a magnetic material is said to have magnetocrystalline anisotropy if it takes more energy to magnetize it in certain directions than in others. $K_\text{a}$ is the magnetocrystalline anisotropy constant and $\mathbf{u}$ is the unit vector along the anisotropy direction. For Fe-Ni Permalloy, $K_\text{a}$ is usually taken as zero. The demagnetization energy
\begin{equation}
E^\text{demag}=-\frac{\mu_0}{2}M_\text{s}\mathbf{m}\mathbf{\cdot}\mathbf{H}_\text{d}
\label{eqdemag}
\end{equation}
in which $\mu_0$ is the vacuum permeability constant and $\mathbf{H}_\text{d}$ is the demagnetization field generated by the magnetic body itself. The Zeeman energy
\begin{equation}
E^\text{Zeeman}=-\mu_0M_\text{s}\mathbf{m}\mathbf{\cdot}\mathbf{H}_\text{ex}
\label{eqzeeman}
\end{equation}
is the energy of magnetization in an external magnetic field $\mathbf{H}_\text{ex}$.

All these above energy terms are only related to the magnetization. The most important term considered here is $E^\text{m-ela}$, i.e. the magnetoelastic energy originated from the coupling between magnetization and stress/strain, which for a polycrystalline with isotropic magnetostriction can be given as \cite{manfred2003micromagnetism}
\begin{equation}
E^\text{m-ela}=-\frac{3}{2}\lambda_\text{s}\left(\sum_{i=1}^3\sigma_{ii}m_i^2 + \sum_{i\neq j}^3\sigma_{ij}m_i m_j \right),
\label{eq13}
\end{equation}
where $\lambda_\text{s}$ is the magnetostriction coefficient. It can be seen from Eq. (\ref{eq13}) that the residual stress $\sigma_{ij}$ induced by the SLM process contributes to the total free energy and thus will affect the magnetic hysteresis and coercivity.

\begin{algorithm*}{!t}
\caption{Conjugate gradient method for energy minimization and hysteresis calculation in micromagnetics}
\begin{algorithmic}[1]
\State $i \gets 0$
\State $\hat{\mathbf{m}}^{(0)} \gets$ initial magnetization direction
\While {$i \leq 2N$}
\State $\hat{\mathbf{H}}_\text{ex}^{(i)} \gets 2\hat{\mathbf{H}}_\text{ex}^\text{max}\|i/N-1\|-\hat{\mathbf{H}}_\text{ex}^\text{max}$
\State Minimize $E^\text{tot}(\hat{\mathbf{m}},\hat{\mathbf{H}}_\text{ex}^{(i)})=\frac{1}{2}\hat{\mathbf{m}}^\text{T}\mathbf{C}\hat{\mathbf{m}}-\frac{1}{2}\hat{\mathbf{H}}_\text{d}^\text{T}\hat{\mathbf{m}}-\hat{\mathbf{H}}_\text{ex}^{(i)\text{T}}\hat{\mathbf{m}}$:
      \State $j \gets 0$
      \State Initial magnetization direction at $\hat{\mathbf{H}}_\text{ex}^{(i)}$: $\hat{\mathbf{m}}_0, \mathbf{m}_0^I \gets \hat{\mathbf{m}}^{(i)}$
      \State Effective magnetic field at cell $I$: $\mathbf{H}_0^I \gets -\frac{1}{\mu_0 M_\text{s}^I}\nabla_{\mathbf{m}_0^I} E^\text{tot}(\hat{\mathbf{m}}_0,\hat{\mathbf{H}}_\text{ex}^{(i)})$
      \State Initial search direction: $\mathbf{d}_0 \gets \widehat{\pmb{\mathbb{H}}}_0= [(\mathbf{m}_0^1\times\mathbf{H}_0^1\times\mathbf{m}_0^1)^\text{T},\dots,(\mathbf{m}_0^K\times\mathbf{H}_0^K\times\mathbf{m}_0^K)^\text{T}]^\text{T}$
       \While {$\| \widehat{\pmb{\mathbb{H}}}_j \| > \wp$}
       \State $\alpha_j \gets \text{minimizing}$ $E^\text{tot}(\hat{\mathbf{m}}_j+\alpha_j \mathbf{d}_j ,\hat{\mathbf{H}}_\text{ex}^{(i)})$ w.r.t. $\alpha_j$ (line search)
       \State $\hat{\mathbf{m}}_{j+1} \gets \hat{\mathbf{m}}_j + \alpha_j \mathbf{d}_j$
       \State $\hat{\mathbf{m}}_{j+1}, \mathbf{m}_{j+1}^I  \gets \hat{\underline{\mathbf{m}}}_{j+1}$ (renormalized by Eq. (\ref{eqnorm}))
       \State $\mathbf{H}_{j+1}^I \gets -\frac{1}{\mu_0 M_\text{s}^I}\nabla_{\mathbf{m}_{j+1}^I} E^\text{tot}(\hat{\mathbf{m}}_{j+1},\hat{\mathbf{H}}_\text{ex}^{(i)})$
       \State $\widehat{\pmb{\mathbb{H}}}_{j+1} \gets [(\mathbf{m}_{j+1}^1\times\mathbf{H}_{j+1}^1\times\mathbf{m}_{j+1}^1)^\text{T},\dots,(\mathbf{m}_{j+1}^K\times\mathbf{H}_{j+1}^K\times\mathbf{m}_{j+1}^K)^\text{T}]^\text{T}$
       \State $   \beta \gets
\left\{
             \begin{array}{lr}
            \widehat{\pmb{\mathbb{H}}}_{j+1}^\text{T}\widehat{\pmb{\mathbb{H}}}_{j+1}/ (\widehat{\pmb{\mathbb{H}}}_j^\text{T}\widehat{\pmb{\mathbb{H}}}_j ) & \text{(Fletcher--Reeves method \cite{fletcher1964function})} \\
            \widehat{\pmb{\mathbb{H}}}_{j+1}^\text{T} (\widehat{\pmb{\mathbb{H}}}_{j+1}- \widehat{\pmb{\mathbb{H}}}_j ) / (\widehat{\pmb{\mathbb{H}}}_j^\text{T}\widehat{\pmb{\mathbb{H}}}_j ) & \text{(Polak--Ribiere method \cite{polak1969note})}
             \end{array}
\right. $
       \State $\mathbf{d}_{j+1} \gets \widehat{\pmb{\mathbb{H}}}_{j+1} + \beta \mathbf{d}_j$
       \State $j \gets j+1$
       \EndWhile
       \State Final magnetization direction at $\hat{\mathbf{H}}_\text{ex}^{(i)}$: $\hat{\mathbf{m}}^{(i)} \gets \hat{\mathbf{m}}_j$
\EndWhile
\end{algorithmic}
\label{algorithm1} 
\end{algorithm*}

\begin{figure*}[!b]
\centering
\includegraphics[width=12cm]{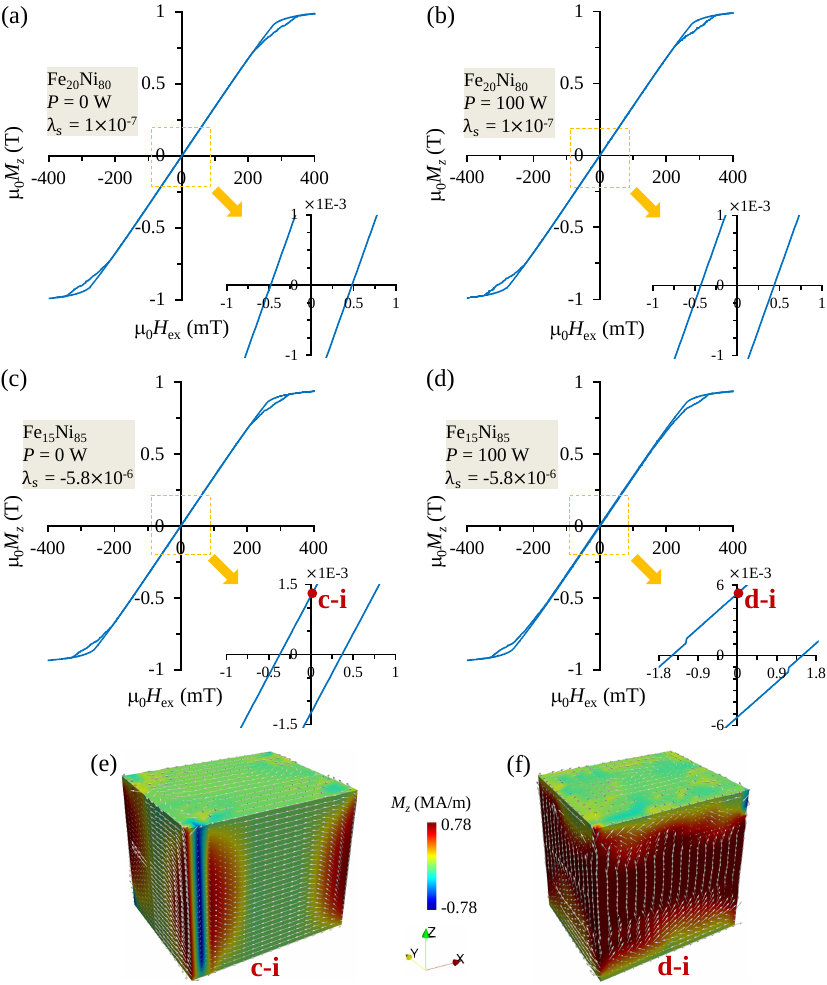}
\caption{Magnetic hysteresis predicted by micromagnetic simulations. (a) Non-processed and (b) SLM processed Fe$_{20}$Ni$_{80}$. (c) Non-processed and (d) SLM processed Fe$_{15}$Ni$_{85}$. Remanent magnetic configuration: (e) point c-i in (c) and (f) point d-i in (d).}
\label{fo1}
\end{figure*}

The minimization of total energy $E^\text{tot}$ with respect to $\mathbf{m}$ can be realized by the conjugate gradient method. The conjugate gradient method for the hysteresis calculation is given in Algorithm \ref{algorithm1}, which is implemented in OOMMF code \cite{donahue1999oommf} in the framework of finite difference method (FDM). The details of the numerical implementation are concisely described here. 
Through the discretization by FDM, the energy in Eq. (\ref{eqEtot}) at an external magnetic field $\mathbf{H}_\text{ex}$ can be rewritten as
\begin{equation}
E^\text{tot}(\hat{\mathbf{m}},\hat{\mathbf{H}}_\text{ex})=\frac{1}{2}\hat{\mathbf{m}}^\text{T}\mathcal{C} \hat{\mathbf{m}}-\frac{1}{2}\hat{\mathbf{H}}_\text{d}^\text{T}\mathcal{M} \hat{\mathbf{m}}-\hat{\mathbf{H}}_\text{ex}^\text{T}\mathcal{M} \hat{\mathbf{m}}.
\label{eqEtot_d}
\end{equation}
In Eq. (\ref{eqEtot_d}), the magnetization direction vectors $\mathbf{m}^I$ at each cell $I$ of a FDM mesh are gathered into a vector $\hat{\mathbf{m}} \in \mathbb{R}^{3K}$ ($K$ is the total number of FDM cells), i.e. 
\begin{equation}
\hat{\mathbf{m}}=\left[ m_x^1,m_y^1,m_z^1,\dots,m_x^K,m_y^K,m_z^K \right]^\text{T}.
\label{eqhatm}
\end{equation}
In the similar way, the external magnetic field and the demagnetizing field at the nodes or cells are gathered into $\hat{\mathbf{H}}_\text{ex}$ and $\hat{\mathbf{H}}_\text{d}$, respectively. 
Since the magnetization magnitude does not change and only the magnetization direction varies (i.e. $\|\mathbf{m}^I\|=1$), a renormalized vector $\hat{\underline{\mathbf{m}}}$ has to be used during the numerical calculations and is correspondingly defined as
\begin{equation}
\hat{\underline{ \mathbf{m}}}=\left[\frac{m_x^1}{\|\mathbf{m}^1\|},\frac{m_y^1}{\|\mathbf{m}^1\|},\frac{m_z^1}{\|\mathbf{m}^1\|},\dots,\frac{m_x^K}{\|\mathbf{m}^K\|},\frac{m_y^K}{\|\mathbf{m}^K\|},\frac{m_z^K}{\|\mathbf{m}^K\|} \right]^\text{T}.
\label{eqnorm}
\end{equation}
The sparse matrix $\mathcal{C}$ contains grid information associated with the exchange, anisotropy, and magnetoelastic energies. The matrix $\mathcal{M}$ accounts for the local variation of the saturation magnetization $M_\text{s}$ within the magnet. The effective magnetic field at each cell $I$ is calculated as
\begin{equation}
\mathbf{H}^I=-\frac{1}{\mu_0M_\text{s}^I}\nabla_{\mathbf{m}^I}E^\text{tot}(\hat{\mathbf{m}},\hat{\mathbf{H}}_\text{ex}).
\label{eqHeff}
\end{equation}
It should be mentioned that in contrast to the normal conjugate gradient method which directly includes the energy gradient (e.g. $\mathbf{H}^I$ in Eq. (\ref{eqHeff})) in the calculation of search direction, here the cross product $\pmb{\mathbb{H}}^I=\mathbf{m}^I\times\mathbf{H}^I\times\mathbf{m}^I$ is used, as shown in Algorithm \ref{algorithm1}. By using the constraint $\|\mathbf{m}^I\|=1$, $\pmb{\mathbb{H}}^I$ is simplified as $\mathbf{H}^I-(\mathbf{H}^I\cdot\mathbf{m}^I)\mathbf{m}^I$, i.e. it only represents the effective magnetic field component perpendicular to $\mathbf{m}^I$. The application of this cross product is physically reasonable, since the field parallel to $\mathbf{m}^I$ cannot induce the magnetization change according to the Landau--Lifshitz--Gilbert equation \cite{gilbert2004phenomenological,yi2014constraint,yi2016micromagnetic,yi2017multiscale}. Following the similar notation for $\hat{\mathbf{m}}$, this cross product vectors $\pmb{\mathbb{H}}^I$ at each cell $I$ of a FDM mesh are gathered into a vector $\widehat{\pmb{\mathbb{H}}} \in \mathbb{R}^{3K}$, i.e.
\begin{equation}
\widehat{\pmb{\mathbb{H}}} = [(\mathbf{m}^1\times\mathbf{H}^1\times\mathbf{m}^1)^\text{T},\dots,(\mathbf{m}^K\times\mathbf{H}^K\times\mathbf{m}^K)^\text{T}]^\text{T}.
\label{eqmHm}
\end{equation}
In order to compute the value of $\beta$ in Algorithm \ref{algorithm1} for the update of the search direction in the conjugate gradient method, both the Fletcher--Reeves method \cite{fletcher1964function} and Polak--Ribiere method \cite{polak1969note} are numerically implemented. It should be mentioned that for the Polak--Ribiere method, the vector $\widehat{\pmb{\mathbb{H}}}_j$ is needed. While for the Fletcher--Reeves method, only the scalar values $\widehat{\pmb{\mathbb{H}}}_j^\text{T}\widehat{\pmb{\mathbb{H}}}_j$ needs to be saved between iterations, which reduces the memory requirement and thus is chosen for the micromagnetic energy minimization in this work. Repeating the minimization under different external magnetic fields $\hat{\mathbf{H}}_\text{ex}^{(i)}$ to give the corresponding $\hat{\mathbf{m}}^{(i)}$ (Algorithm \ref{algorithm1}) will result in the magnetic hysteresis from which the magnetic properties can be calculated.

The magnetic parameters of Fe-Ni alloy with three different compositions used for micromagnetic simulations are listed in Table \ref{tab1} \cite{bonin2005dependence}. The nominal magnetostriction of Fe$_{20}$Ni$_{80}$ is zero. In order to investigate the possible effect of tiny deviation from the nominally zero magnetostriction, small $\lambda_\text{s}$ ($\pm 1.0\times 10^{-7}$) is also considered for the case of Fe$_{20}$Ni$_{80}$. Fe$_{15}$Ni$_{85}$ and Fe$_{25}$Ni$_{75}$ possess negative and positive magnetostriction, respectively. The micromagnetic model is meshed by cubic cells with a size of  10 $\mu$m $\times$ 10 $\mu$m $\times$ 5 $\mu$m.  Since the mesh for stress calculation and micromagnetic simulation is different, the stress distribution in the micromagnetic model is obtained by performing interpolation of the nodal stress from the previous mechanical analysis through a Delaunay triangulation of the scattered data \cite{amidror2002scattered}.

\newcommand{\tabincell}[2]{\begin{tabular}{@{}#1@{}}#2\end{tabular}}
\begin{table}
\centering 
\caption{Magnetic parameters \cite{bonin2005dependence} of Fe-Ni alloy for micromagnetic simulations}
\begin{tabular}{cccc}
\hline
composition & $\lambda_\text{s}$ & $M_\text{s}$ (MA/m) & $A_\text{e}$ (pJ/m)\\ \hline
Fe$_{25}$Ni$_{75}$ & $7.3\times 10^{-6}$ & 0.89 & 13 \\
Fe$_{20}$Ni$_{80}$ & \tabincell{c}{$1.0\times 10^{-7}$ \\ 0 \\  $-1.0\times 10^{-7}$} & 0.83 & 13 \\ 
Fe$_{15}$Ni$_{85}$ & $-5.8\times 10^{-6}$ & 0.78  & 13\\
    \hline
\end{tabular}
\label{tab1}
\end{table}

\begin{figure*}[!t]
\centering
\includegraphics[width=12cm]{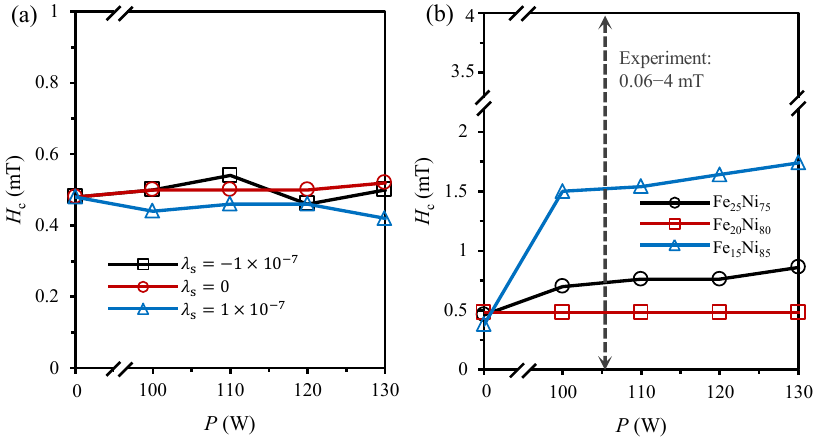}
\caption{Calculated magnetic coercivity of SLM processed Fe-Ni alloy under different laser power. (a) Coercivity of SLM processed Fe$_{20}$Ni$_{80}$ with different magnetoelastic coefficient. (b) Coercivity of SLM processed Fe$_{25}$Ni$_{75}$, Fe$_{20}$Ni$_{80}$, and Fe$_{15}$Ni$_{85}$. The arrow in (b) indicates the experimentally measured coercivity ranging form 0.06 to 4 mT \cite{zhang2012studies,zhang2013microstructure,zhang2013magnetic}.}
\label{fo2}
\end{figure*}

The magnetic hysteresis is calculated by applying external magnetic field along the $z$ direction. Fig. \ref{fo1}(a)-(d) shows typical hysteresis curves for the case of Fe$_{20}$Ni$_{80}$ and Fe$_{15}$Ni$_{85}$ with and without SLM processing. Zooming in around the original point leads to the coercivity observable. When there is no SLM processing, both Fe$_{20}$Ni$_{80}$ and Fe$_{15}$Ni$_{85}$ possess very small coercivity around 0.5 mT, as depicted in Fig. \ref{fo1}(a) and (c). This is expected for the magnetically soft Fe-Ni alloys. For Fe$_{20}$Ni$_{80}$, we check the influence of small magnetostriction coefficient on its coercivity, as shown in Fig. \ref{fo1}(b) and Fig. \ref{fo2}(a). It can be seen that if $\lambda_\text{s}$ is set as small values $\pm 1.0\times 10^{-7}$, SLM processing with different laser power can only lead to a coercivity change within 0.05 mT. On the contrary, for Fe$_{15}$Ni$_{85}$ which has a negative $\lambda_\text{s}$ of $-5.8\times 10^{-6}$, SLM processing with 100 W laser can increase the coercivity to $\sim 1.5$ mT, as presented in Fig. \ref{fo1}(d).
Fig. \ref{fo2}(b) shows the calculated coercivity for different alloy compositions and laser power. It can be found that the coercivity of SLM processed Fe$_{15}$Ni$_{85}$ and Fe$_{25}$Ni$_{75}$ is increased to $\sim$1.7 and $\sim$0.8 mT, respectively. Within the laser power $P=100-130$ W, the coercivity is found to only slightly increase with $P$. The reason may be that the stress magnitude and distribution are not significantly changed within $P=100-120$ W. It should be mentioned that the experimentally measured coercivity of SLM processed Fe-Ni alloys is around $0.06-4$ mT \cite{zhang2012studies,zhang2013microstructure,zhang2013magnetic}. Our simulation results on the coercivity are in accordance with these experimental measurement.
Apart from the coercivity values, the remanent magnetization ($\mu_0 M_\text{r}$) and magnetic domain structure are also affected by the SLM process.
Fig. \ref{fo1}(e) and (f) shows the magnetic configuration at the remanent state of the sample in (c) and (d), respectively.
For Fe$_{20}$Ni$_{80}$ (Fig. \ref{fo1}(a) and (b)) and Fe$_{15}$Ni$_{85}$ without SLM processing (Fig. \ref{fo1}(c)), $\mu_0 M_\text{r}$ is as low as $\sim$1 mT. Accordingly, the remanent magnetic configuration is composed of large-area magnetic domains perpendicular to $z$ direction or along the negative $z$ direction, as shown in Fig. \ref{fo1}(e).
However, after Fe$_{15}$Ni$_{85}$ is processed by SLM, $\mu_0 M_\text{r}$ is enhanced to $\sim$5 mT (Fig. \ref{fo1}(d)), and magnetic domains along the positive $z$ direction occupy larger areas (Fig. Fig. \ref{fo1}(f)). These results computationally confirm that SLM processing could affect the coercivity, remanent magnetization, and magnetic domain structure in Fe-Ni alloys.
\begin{figure*}[!b]
\centering
\includegraphics[width=14cm]{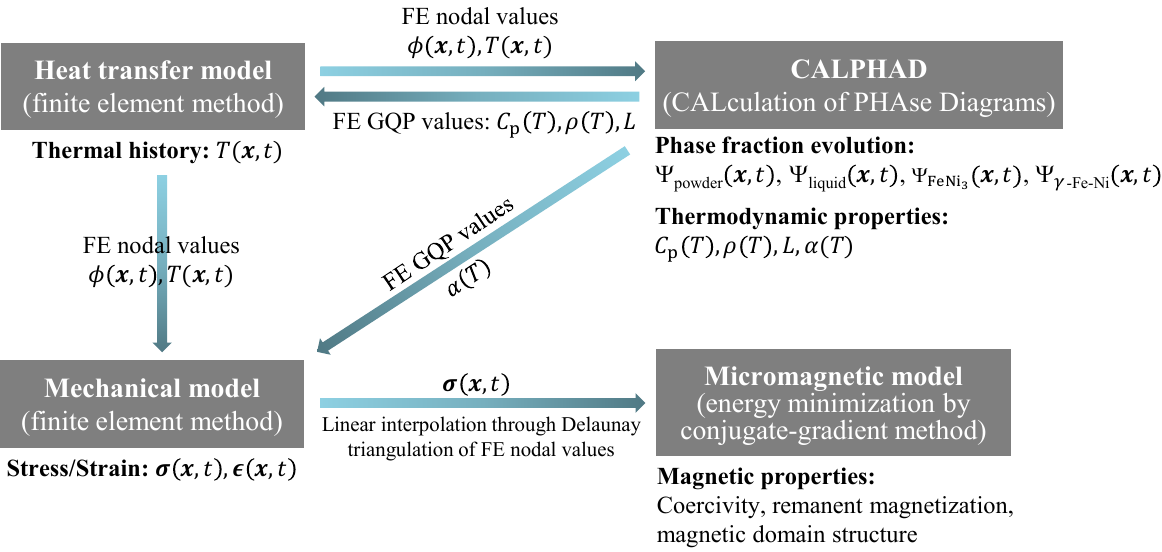}
\caption{Flowchart and coupling schemes of the integrated framework for the computational study of laser additively manufactured magnetic materials. FE: finite element; GQP: Gaussian quadrature point.}
\label{figFlowChart}
\end{figure*}

\section{Summary and outlook}
In conclusion, we have integrated FEA, CALPHAD output, and micromagnetics to demonstrate the first attempt for the computational evaluation of the microstructure evolution and magnetic coercivity of SLM processed magnetic Fe-Ni alloys.
The flowchart and coupling schemes of the integrated framework are summarized in Fig. \ref{figFlowChart}, including heat-transfer model, mechanical model, CALPHAD, and micromagnetic model. Both the heat-transfer and mechanical models are numerically solved by finite element method. The micromagnetic model and thus the magnetic hysteresis are calculated through the energy minimization by the conjugate gradient method within the finite difference framework. These models are coupled through the information transfer among them. For example, the temperature history $T(\bm{x},t)$ from the heat transfer model can be used as input for the CALPHAD and mechanical model. In detail, finite element (FE) nodal values of field variable $\phi(\bm{x},t)$ and temperature $T(\bm{x},t)$ from the finite element simulation of heat transfer model are mapped to the CALPHAD model by a Python script to obtain the temporal and spatial phase fraction. Meanwhile, FE nodal values of $\phi(\bm{x},t)$ and $T(\bm{x},t)$ are also imported to the mechanical model for the calculation of transient stress $\pmb{\sigma}(\bm{x},t)$ and strain $\pmb{\epsilon}(\bm{x},t)$ by finite element method. In turn, CALPHAD can provide thermodynamically consistent parameters for the Gauss quadrature points in finite element simulations, such as specific heat $C_\text{p}(T)$, density $\rho(T)$, and latent heat $L$ for the heat transfer model and thermal expansion coefficient $\alpha(T)$ for the mechanical model. Furthermore, the stress fields from the mechanical model can be input to the micromagnetic model for the calculation of magnetic properties, including coercivity, remanent magnetization, and magnetic domain structure. For transferring the stress fields from the finite element method to the micromagnetic model which is solved by finite difference method, we use a Delaunay triangulation of the scattered FE nodal values to perform the linear interpolation. Overall, by using the temperature-dependent material states/parameters and the step-by-step element activation of powder mesh, finite element simulation of the heat-transfer model is performed to calculate the temperature distribution and evolution during the SLM processing of Fe-Ni alloys.
With the thermal history from finite element simulations as input, thermomechanical analysis is performed by using the elastic-plastic material constitutive model.  By integrating the thermal history and CALPHAD output, the evolution and distribution of liquid, powder, FeNi$_3$ and $\gamma$-Fe-Ni during the SLM process are calculated. 
Finally, micromagnetic simulations which treats the residual stress as the magnetoelastic energy are carried out to calculate the magnetic property of SLM processed Fe-Ni alloy.
By using this computational framework, the melting pool geometry and the cyclic thermal history are identified. The cyclic tension and compression are confirmed in the interface of two neighboring tracks, which could degrade the interface bonding. It is found that the material firstly changes from powder to liquid and then experiences cyclic phase changes between $\gamma$-Fe-Ni and FeNi$_3$. SLM process is found to obviously enhance the remanent magnetization and increase the coercivity of Fe$_{15}$Ni$_{85}$ and Fe$_{25}$Ni$_{75}$ to $0.8-1.7$ mT. The calculated coercivity is shown to agree with the experimental values.

While we have shown a promising method for the simulation of additively manufactured magnetic materials by integrating FEA, CALPHAD, and micromagnetics, the work here is a first attempt and lots of issues have to be thoroughly considered in the near further. As an initial work, here we limit ourselves to the working principle and potential feasibility of the proposed computational scheme, and do not focus on the accurate prediction of industrial or real AM process at the current stage. Several issues in this work related to the real AM process have to be deliberated and resolved in the next step, as listed in the following:

(1) The AM processability of FeNi alloys by SLM is an open question. Here we hypothesize defect-free processing in our simulation. The low processability related effects from fluid dynamics, surface tension and gas flow have to be considered, with more experimental information on the processability.

(2) The temperature-dependent material parameters in Fig. \ref{f2}, which are important for the thermal and stress analysis, are not accurate, due to the lack of sufficient experimental data. It will be good to make more efforts on experimental measurements and obtain thermodynamic property information of magnetic alloys form CALPHAD database. Including the possible vaporization may be also necessary for the accurate prediction of the peak temperature. The phase change and composition variation induced stress should also play a role.

(3) Apart from the weak microsegregation of elements by using the Scheil-Gulliver model, the phase fraction is taken as the main indicator for the microstructure evolution in this work. However, in the real experimental case, microstructure would be more complicated. Other phases except for $\gamma$-Fe-Ni and FeNi$_3$ may also exist and spatial element microsegregation and composition inhomogeneity can occur. There are also other general issues in the prediction of  microstructure evolution during SLM process, such as the non-equilibrium state, fast cooling induced solute trapping, effect of free energy from magnetic contribution, etc. Phase-field simulation using CALPHAD information of magnetic materials \cite{xiong2011magnetic} could be a viable methodology for predicting the microstructure evolution during the SLM processing, and deserves our future efforts.

(4) The dependence of magnetic properties on the phase fraction and microstructure is very complicated and not involved in this work.
As for the prediction of magnetic properties of SLM-processed magnetic materials in the real case, more microstructure information (e.g. phase distribution, grain boundaries, grain orientation, porosity, surface roughness, etc.) except for residual stress should be considered.

(5) Even though the calculated coercivity is shown to be in line with the experimental one for the laser additively manufactured magnetic FeNi, the simulation work here lacks the experimental validation on several points such melt pool size, temperature, stress, microstructure, etc. Collaborative experimental work has to be carried out in order to make the computational framework fully convinced.
\\[1em]

\noindent
\textbf{Acknowledgements} The support from the German Science Foundation (DFG YI 165/1-1 and DFG XU 121/7-1), the Profile Area From Material to Product Innovation -- PMP (TU Darmstadt), the European Research Council (ERC) under the European Union´s Horizon 2020 research and innovation programme (grant agreement No 743116), and the LOEWE research cluster RESPONSE (Hessen, Germany) is acknowledged. The authors also greatly appreciate their access to the Lichtenberg High Performance Computer of Technische Universit\"at Darmstadt.

\bibliographystyle{cm-num}
\bibliography{mybibfile}

\end{document}